\documentclass[12pt]{article}
\usepackage[margin=1in]{geometry}
\RequirePackage{amsthm,amsmath,amsfonts,amssymb}
\RequirePackage[authoryear]{natbib}
\RequirePackage{graphicx}
\usepackage{url,placeins,bm}
\usepackage{lmodern}
\usepackage{xcolor}
\usepackage{verbatim}
\usepackage{booktabs}
\usepackage{ragged2e}
\usepackage[group-separator={,}]{siunitx}
\newcommand{\bignumber}[1]{\num[group-separator={,}]{#1}}

\newcommand{\sign}[1]{\text{sign}\left(#1\right)}
\newcommand{\drop}[1]{{#1}_{-i}}
\newcommand{\est}[1]{\hat{#1}}
\newcommand{\inv}[1]{{#1}^{-1}}
\newcommand{\indep}{\rotatebox[origin=c]{90}{$\models$}}

\newcommand{\scl}{s^l}
\newcommand{\scs}{s^s}
\newcommand{\scli}{s^l_i}
\newcommand{\scsi}{s^s_i}
\newcommand{\wl}{w^l}
\newcommand{\ws}{w^s}
\newcommand{\wc}{w}
\usepackage[ruled, linesnumbered]{algorithm2e}
\usepackage{algpseudocode}
\usepackage{etoolbox}
\usepackage{mathtools}
\DeclarePairedDelimiter\floor{\lfloor}{\rfloor}

\title{Separating and reintegrating latent variables to improve classification of genomic data}
\author{
Nora Yujia Payne\\
University of Michigan\\
\texttt{yujiap@umich.edu}
\and
Johann A. Gagnon-Bartsch\\
University of Michigan\\
\texttt{johanngb@umich.edu}
  \and
}

\begin{document}

\maketitle

\begin{abstract}
Genomic datasets contain the effects of various unobserved biological variables in addition to the variable of primary interest.  These latent variables often affect a large number of features (e.g., genes) and thus give rise to dense latent variation, which presents both challenges and opportunities for classification.  Some of these latent variables may be partially correlated with the phenotype of interest and therefore helpful, while others may be uncorrelated and thus merely contribute additional noise.  Moreover, whether potentially helpful or not, these latent variables may obscure weaker effects that impact only a small number of features but more directly capture the signal of primary interest.  We propose the cross-residualization classifier to better account for the latent variables in genomic data. Through an adjustment and ensemble procedure, the cross-residualization classifier essentially estimates the latent variables and residualizes out their effects, trains a classifier on the residuals, and then re-integrates the the latent variables in a final ensemble classifier.  Thus, the latent variables are accounted for without discarding any potentially predictive information that they may contribute.  We apply the method to simulated data as well as a variety of genomic datasets from multiple platforms. In general, we find that the cross-residualization classifier performs well relative to existing classifiers and sometimes offers substantial gains.
\end{abstract}

\section{Introduction}
\label{sec:intro}

High-dimensional classification is a ubiquitous and challenging problem in genomics. Classical methods, such as linear discriminant analysis, cannot be used directly since usual estimates for population parameters are poor when the number of features exceeds the number of observations \citep{peck1982use}. For instance, it is well known that the sample covariance matrix is no longer invertible and some form of regularization is necessary for its estimation. Additionally, even when individual parameters can be estimated well, the aggregation of small estimation errors can degrade the performance of a classifier \citep{fan2008high}. Various strategies have been developed to address these challenges. In genomics, popular approaches include various forms of dimension reduction, feature selection \citep{guyon2003introduction,saeys2007review}, as well as the ``independence rule'' \citep{bickel2004some,dudoit2003classification}. These strategies are embedded in many popular classifiers \citep{witten2011penalized,zou2005elasticnet,nguyen2002tumor,fan2008high,tibshirani2002diagnosis,zhang2012classification}.

Many of these approaches work by discarding or disregarding aspects of the data.  For example, feature selection methods aim to select a small subset of relevant features while ignoring many null features. Dimension reduction strategies such as principal components analysis (PCA) or partial least squares (PLS) focus attention on a low-dimensional subspace of the original feature space, disregarding uninformative dimensions that merely contain noise.  The independence rule, which disregards correlations between features, is yet another example. When the assumptions underlying these approaches hold, they can yield substantial improvements in classifier fit and accuracy by reducing the number of parameters one must estimate.

However, these assumptions are not always appropriate for genomic data \citep{hall2014feature}. Consider a hypothetical dataset of methylation signatures generated from a sample of lung tumors. The tumors are of either Type A or Type B, and we wish to train a classifier on the methylation data for the purpose of classifying new lung tumors. Although tumor type is the signal of primary interest in this classification task, it is likely only one of many signals present in the data, many of which arise from various biological and environmental factors. For instance, it has been observed that smokers and nonsmokers differ in their DNA methylation signatures across the whole genome \citep{bollepalli2019epismoker,elliott2014differences,lee2013cigarette,zeilinger2013tobacco,wan2012cigarette}. Air pollution and exposure to fine particulates have also been found to affect gene expression and methylation patterns \citep{bind2014air,quay1998air}. Since these variables affect genetic activity across many genes, the data contain widespread correlations across the features and many dense signals in addition to the signal of primary interest \citep{boyle2017expanded}. As an added complication, these additional variables are generally latent. For instance, smoking status and fine particulate exposure are frequently not recorded, difficult to measure, or underreported. These issues are especially challenging when the signal of primary interest is sparse.

In this setting, the aforementioned strategies for coping with high-dimensionality may not be optimal. Dimension reduction strategies may ignore informative dimensions of the feature space, and feature selection strategies may discard discriminative variables. The independence rule is similarly suboptimal. For example, in \citet{fan2012road}, it is shown that leveraging correlations between variables can yield further reduction in misclassification error, as opposed to simply ignoring them. Thus when dense latent signals are present, commonly used strategies may not only discard noise, but relevant information as well.

There is a growing body of work on high-dimensional estimation and prediction in the presence of dense signals \citep{dobriban2018high,dicker2012optimal,cook2012estimating}, and a substantial portion of this literature focuses on such problems as arising from latent variables specifically \citep{kneip2011factor,zheng2017nonsparse,dicker2012optimal}. One way to account for latent variables is by performing PCA and using the leading principal components as additional predictors in the model; \citet{kneip2011factor} discuss this idea in the context of linear regression. Another approach to account for the effects of latent variables is to assume ``conditional sparsity'' of the coefficient vector or covariance matrix in a linear regression model \citep{zheng2017nonsparse,fan2013large}. Roughly speaking, these conditional methods model and condition on the presence of latent factors in order to accurately estimate the model parameters of interest. 

Our main contribution in this paper is a framework for training a classifier in the presence of latent variables. We focus primarily on the setting in which the signal of primary interest is sparse, the latent signals are dense, and the latent signals are potentially correlated with the signal of primary interest. The key idea is to decompose the data into a dense low-rank component and a sparse component, train separate classifiers on these components, and combine these classifiers into a single ensemble. Importantly, the sparse component is adjusted for the dense low-rank component, so that these two components contain distinct information. 
We propose an algorithm for performing this simultaneous decomposition and adjustment, which we call \emph{cross-residualization}. We also propose a specific instantiation of our framework, the \emph{cross-residualization classifier} (CRC) in the setting where the class-conditional distributions are Gaussian. In this setting, simple linear classifiers may be used on each of the separate components.

The paper is organized as follows. In Section \ref{sec:model-motiv} we introduce the model and provide the motivation for our method, discussing the challenges and opportunities provided by the latent variables in more detail. In Section \ref{sec:cnc} we describe the algorithm for fitting the cross-residualization classifier. Section \ref{sec:simulation} details several simulations which demonstrate the advantages of our approach and offer comparisons to classifiers commonly used in genomic applications. In Section \ref{sec:real-data} we compare our classifier to several commonly used classifiers on a diverse collection of genomic datasets. Section \ref{sec:discussion} concludes.

%%%%%%%%%%%%%%%%%%%%%%%%%%%%%%%%%%%%%%%%%%%%%%
%%%% Section 2
%%%%%%%%%%%%%%%%%%%%%%%%%%%%%%%%%%%%%%%%%%%%%%

\section{Model and Motivation}
\label{sec:model-motiv}

Here we present a sequence of models to illustrate the challenges and opportunities that latent variables present for classification of genomic data. In the first model, latent variables are absent. The second model includes latent variables, but they are uncorrelated with the variable of primary interest. In the third model, latent variables are potentially correlated with the variable of primary interest. We ultimately work under the final model, although the initial models serve to illustrate several key points.

\subsection{Simple model}

Suppose an observation is given by $(S, T)$, where $S \in \mathbb{R}^{1 \times p}$ is the feature vector and $T \in \{ \pm 1 \}$ is the class label. One might imagine $S$ to be a vector of expression levels for $p$ genes and $T$ to be disease status, indicating the presence or absence of disease. Consider a model for $S$ in which no latent factors are present,
\begin{equation} \label{eq:simple-model}
    S_{1 \times p} = T_{1 \times 1}\gamma_{1 \times p} + \epsilon_{1 \times p}
\end{equation}
where $T \in \{\pm 1\}$ with $P(T=1) = \pi \in (0,1)$. We assume that the entries of $\epsilon$ are independent of one another, although their variances are permitted to differ. In particular, we assume $\epsilon$ follows $N(0, \Sigma)$ where $\Sigma = \text{diag}(\sigma_1^2, \ldots, \sigma_p^2)$. For simplicity, we assume that $T$ is binary and $\pi = 1/2$ in this paper. However, our discussion and methodology readily generalize to the multi-class setting and to cases in which the prior probabilities are not uniform.

Under model \eqref{eq:simple-model},
\[
    S \mid T  \sim N(T\gamma, \Sigma).
\]
The optimal Bayes classifier is $f(x) = \sign{w_\text{simple} \cdot x}$, where
\[
	w_\text{simple} \propto \inv{\Sigma}\gamma' = (\gamma_1 / \sigma_1^2, \ldots, \gamma_p / \sigma_p^2)'
\]
and where $\sign{\cdot}$ denotes the sign function. In many applications of practical interest, the vector $\gamma$ is sparse since the variable of primary interest affects only a small fraction of the genes. In this case, because $\inv{\Sigma}$ is a diagonal matrix, the optimal weight vector $w_\text{simple}$ is also sparse. As a result, the independence rule and variable selection techniques can be used to efficiently estimate $w_\text{simple}$, despite it being a $p$-dimensional parameter.

We make a brief remark regarding the notation in \eqref{eq:simple-model}. In a typical classification model, we might ordinarily denote the feature vector $S$ by `$X$' and the class label $T$ by `$Y$'. However, we regard $S$ as the response in a generative model for the features, so it appears on the left hand side of \eqref{eq:simple-model} while the class labels $T$ appear on the right hand side. As a result, using `$X$' and `$Y$' may cause confusion. We use the more neutral $S$ and $T$, the latter of which can be thought of as representing the `target' in a classification analysis.

\subsection{Uncorrelated latent variables}

In reality, genomic data contain the effects of biological variables other than the variable of primary interest. These variables are generally latent, so observations might be more appropriately modeled via a latent factor model
\begin{equation} \label{eq:lat-fac-model}
	\begin{aligned}
		Z_{1 \times p} 
		&= T_{1 \times 1}\gamma_{1 \times p} + L_{1 \times r}\alpha_{r \times p} + \epsilon_{1 \times p} %\\
		%&= S + \Gamma,
	\end{aligned}
\end{equation}
where $L$ represents the latent variables. Here we assume that $L$ follows $N(0, \Psi)$ and is uncorrelated with $T$, i.e.
\[
	L \mid T \sim N(0, \Psi).
\]
We assume that the latent biological variables tend to affect a large proportion of genes, so the coefficient matrix $\alpha$ is dense. The number of latent variables, $r$, is unknown, but we assume that $r < n-1$. Note that the feature vector in model \eqref{eq:lat-fac-model} is denoted by $Z$ instead of $S$; we continue to reserve the latter to denote $S=T\gamma + \epsilon$ (which is unobserved in this model).  Similar to before, we can think of the feature vector $Z$ as representing the expression levels of $p$ genes and $T$ as disease status, although gene expression levels are now a function of the latent variables in addition to disease status. Latent variable models similar to \eqref{eq:lat-fac-model} appear frequently in the batch effect normalization literature, where the latent variables represent unwanted technical factors to be removed \citep{leek2007capturing, leek2008general, listgarten2010correction, gagnonbartsch2012using, sun2012multiple, parker2014removing, wang2017confounder}. Here, however, we consider the latent variables to be biological variables responsible for variation across a large number of features.

Under the uncorrelated factor model in \eqref{eq:lat-fac-model},
\begin{equation} \label{eq:Z|T-uncorr}
	Z  \mid  T  \sim N(T\gamma, \alpha'\Psi\alpha + \Sigma).
\end{equation}
The optimal Bayes classifier is again a linear classifier of the form $f(x) = \sign{w_\text{uncorr} \cdot x}$, where
\[
    w_\text{uncorr} \propto \inv{(\alpha'\Psi\alpha + \Sigma)}\gamma'.
\]
Unlike $w_\text{simple}$, the optimal weight vector $w_\text{uncorr}$ is dense, even if $\gamma$ is sparse.  Consequently, strategies such as feature selection and the independence rule, which were effective under the simple model, may no longer be ideal.  

This has implications for classification analyses of genomic data, since latent variables are common and presumably exist in nearly every genomic dataset. For example, in all of the datasets we examine in Section 5, more than half of the variance is captured in the first ten principal components (Table \ref{table:pct_var_explained}). Even when these latent variables are uncorrelated with the variable of interest, their presence will result in an optimal weight vector that is non-sparse. Commonly-used strategies such as sparse feature selection will result in many zero feature weights when they are optimally non-zero.

Under \eqref{eq:lat-fac-model}, we observe $Z$ but not $S$. However, observing $S$ would be preferable to observing $Z$, for the reasons highlighted above.  If it were possible to extract $S$ from $Z$, then classification could be performed on $S$ using usual techniques.

\subsection{Correlated latent variables} \label{correlatedmodel}
Now consider a model in which the latent variables $L$ are correlated with $T$. We continue to model an observation by
\[
	Z_{1 \times p} = T_{1 \times 1}\gamma_{1 \times p} + L_{1 \times r}\alpha_{r \times p} + \epsilon_{1 \times p},
\]
but now let
\[
	L \mid T \sim N(T\eta, \Psi)
\]
where $\eta \in \mathbb{R}^{1 \times r}$. Note that if $\eta = 0$, all $r$ latent variables are uncorrelated with $T$ and this model is equivalent to the uncorrelated latent factor model.

Under the correlated latent factor model,
\begin{equation} \label{eq:Z|T-corr}
     Z  \mid  T \sim N\left( T (\gamma + \eta\alpha), \alpha'\Psi\alpha + \Sigma \right).
\end{equation}
The optimal Bayes classifier is once again linear, but with weights
\begin{equation} \label{eq:w_corr}
	w_\text{corr} \propto \inv{(\alpha'\Psi\alpha + \Sigma)}(\gamma + \eta\alpha)'.
\end{equation}
The correlated latent variable model presents the same challenges as the uncorrelated latent variable model. Since the covariance matrix of the class-conditional distribution is non-diagonal, the vector $w_\text{corr}$ is dense, even if $\gamma$ is sparse.

However, when the latent variables are correlated with $T,$ they can provide valuable discriminative information, as seen in the form of the feature weights in \eqref{eq:w_corr}. The class-conditional means under the correlated latent factor model are $\pm (\gamma + \eta\alpha)$, compared to $\pm \gamma$ under the uncorrelated latent factor model. As a result, there will typically be more separation between the two classes when $L$ is correlated with $T$, and it will be easier to discriminate between the two classes. If $\gamma$ is weak and sparse, as it is in many settings of practical interest, then the presence of latent variables correlated with the class label is especially valuable. Observing $S$ is no longer preferable to observing $Z$ when $L$ and $T$ are correlated, as $S$ no longer contains all the predictive information present in $Z$.

\subsection{Discussion}  \label{sec2discussion}

Latent variables present both challenges and opportunities in a high-dimensional classification analysis. In the presence of dense latent signals, the optimal classifier is defined by a dense, $p$-dimensional vector of weights. Although it is common to rely on strategies such as feature selection and the independence rule to estimate this weight vector, these strategies may discard relevant and potentially discriminative information, particularly in the case where the latent variables are correlated with the class label.

Henceforth we work with the correlated latent factor model, which subsumes the simple model in \eqref{eq:simple-model} and the uncorrelated latent factor model in \eqref{eq:lat-fac-model}. 
If there are no latent factors, then $Z$ is simply $S,$ and we can use the aforementioned strategies to train a classifier. Training a classifier is more complicated, however, when the model includes the $L\alpha$ term. If $L$ and $T$ are uncorrelated, a reasonable strategy would be to first isolate $S$ from $Z$ and then train a classifier on $S.$ Such a strategy is reasonable because $(Z \indep T) \mid S$ when $L$ and $T$ are uncorrelated. In other words, $S$ contains all of the predictive information available in $Z.$ However, this is no longer true when $L$ and $T$ are correlated, since $L$ now also contains predictive information. While recovering $S$ alone is insufficient under the correlated model, recovering both $S$ and $L$ would be, in the sense that $(Z \indep T) \mid S, L$.

This observation suggests a procedure in practice. Suppose for a moment that $L$ were known. We could use $L$ to estimate $\alpha,$ and then residualize the latent variables from $Z$ to obtain an estimate of $S,$ i.e. $\hat{S} = Z - L\hat{\alpha}.$ With both $\hat{S}$ and $L$ in hand, we could train separate classifiers on $S$ and $L$, and then combine them in an ensemble, in the spirit of the super learner \citep{polley2011super}. Of course, $L$ is unobserved under our model and must be estimated in order to carry out this procedure. Fortunately, $L$ can be well-approximated by the top $r$ left singular vectors of $Z$ when $\gamma$ is sparse and $p \gg n$. In fact, in the limit it is possible to recover $L$ perfectly, up to an arbitrary parameterization (see Appendix \ref{recovering_L} for conditions).

Even so, challenges remain. If we train a classifier on $\hat{S}$ and $\est{L},$ then applying this classifier to out-of-sample observations entails estimating analogous quantities for out-of-sample observations. There is also the question of how to estimate $r,$ the rank of $L.$ Additionally, utilizing an estimate of $L$ to estimate $S$ may lead to downstream overfitting in the classifier trained on $\hat{S}.$ The ensemble may suffer as a result. We address these challenges in Section 3, in which we lay out algorithms for training a classifier (the CRC) according to the framework outlined above. 
% We call the resulting classifier the \emph{cross-residualization classifier} (CRC). 

In separating the dense latent signals from the sparse signal of interest, one can view the CRC as performing a form of forward selection. The former can be viewed as a common effect shared across all features and the latter a feature specific effect \citep{kneip2011factor}. By residualizing out $L\alpha$ (or an approximation thereof), we isolate the sparse signals so that we learn which of the $p$ features has an individual effect beyond shared the latent effect.  The meta-classifier utilizes this information to improve classification accuracy beyond what can be achieved by a classifier trained on the dense latent signals alone. Because the classifiers in the ensemble are roughly uncorrelated, the utility of the meta-classifier is enhanced. 

We conclude this section with some notation. Let the training set be an i.i.d. sample from the correlated latent factor model in \eqref{eq:Z|T-corr}. Denote the training observations by $\{(Z_1, T_1), \ldots, (Z_n, T_n)\}$ where $Z_i = T_i \gamma + L_i \alpha + \epsilon_i$ for $i=1, \ldots, n.$ Denote the training data in matrix form by
\begin{align*}
	\bm{Z}_{n \times p} 
	&= [Z_1', \ldots, Z_n']' \\
	&= \bm{T}_{n \times 1}\gamma_{1 \times p} + \bm{L}_{n \times r}\alpha_{r \times p} + \bm{\epsilon}_{n \times p} 
\end{align*}
and define $\drop{\bm{Z}}$ to denote the $(n-1) \times p$ matrix consisting of all rows in $\bm{Z}$ except the $i$th.  Finally, denote a generic \emph{target observation}, an out-of-sample observation for which we wish to make a class prediction, by $Z = T\gamma + L\alpha + \epsilon$.   

Note that we do not include an intercept term in the model.  In practice we may re-center the columns of $\bm{Z}$ to have mean 0 and apply the same re-centering to $Z$ (see Appendix \ref{centering}).

%%%%%%%%%%%%%%%%%%%%%%%%%%%%%%%%%%%%%%%%%%%%%%
%%%% Section 3
%%%%%%%%%%%%%%%%%%%%%%%%%%%%%%%%%%%%%%%%%%%%%%

\section{The cross-residualization classifier} \label{sec:cnc}

The CRC is an ensemble of two linear discriminant-based classifiers, which we refer to as CRC-L and CRC-S. CRC-L and CRC-S are trained on the dense latent signals and the sparse signal of interest, respectively. A meta-classifier is fit to the resulting discriminant scores to form the ensemble.

\subsection{CRC-L: Training a classifier on the latent signals}
\label{subsec:CRC-L}
One strategy for fitting a classifier on the latent signals is to use PC-LDA, principal components linear discriminant analysis. That is, first estimate $\bm{L}$ by projecting $\bm{Z}$ onto the top $r$ principal components of $\bm{Z}$ and then train an LDA classifier on this estimate.  Given that $\bm{L}$ can be recovered by the principal components (Appendix \ref{recovering_L}) and that the class conditional distribution of $\bm{L}$ is Gaussian under our model, PC-LDA is a natural strategy.  One challenge, however, is that $r$ is unknown and must be estimated.  If $\hat{r} < r,$ then we may fail to capture all relevant information in $\bm{L}.$ On the other hand, if $\hat{r} > r,$ then the estimate $\hat{\bm{L}}$ will overfit the training data, potentially leading to downstream overfitting in the ensemble. In general, estimating $r$ is a challenging task \citep{wold1978cross, peres2005many, choi2017selecting, virta2019estimating}.

We take an alternative approach which does not require an estimate of $r$ or even an explicit estimate of $\bm{L}.$ Specifically, we project $\bm{Z}$ onto all principal components of $\bm{Z}$ and then use LDA to train a classifier on the projected data (details in Appendix \ref{crcl_implementation}).  This approach amounts to PC-LDA using all $n$ principal components, which has good performance on out-of-sample observations despite overfitting the training data.  To see this, consider the regression analogue of PC-LDA, principal components regression (PCR).  Principal components regression (PCR) and ridge regression are well-known to be closely related \citep{jolliffe1986principal, friedman2001elements}, and are in fact equivalent in a special case.  If $n$ principal components are used, PCR is equivalent to ridge regression with the ridge penalty tending to zero, a technique more commonly known as ridgeless regression \citep{hastie2019surprises}.  Ridgeless regression interpolates the training data, resulting in perfect overfitting. However, it has been shown to have good out-of-sample predictive accuracy when $p \gg n$ \citep{hastie2019surprises}.  This property motivates our approach. So long as we apply CRC-L to out-of-sample observations, we avoid estimating $r$ without suffering too much from the negative effects of overfitting.  Overfitting in the training data remains an issue when fitting the ensemble, but is easily addressed via a leave-one-out approach (Section \ref{sec:ensemble}).

\emph{Comment}:  In this paper, we focus on the case in which $\gamma$ is sparse, and one might wonder what happens when this assumption is violated.  If $\gamma$ is not sparse and the elements of $\gamma$ are sufficiently large (i.e., above noise level), then the signal of interest will be captured in the principal components of $\bm{Z}.$ CRC-L will no longer be a classifier trained strictly on the latent signals, but rather the latent signals and the signal of interest.  In this case, the inclusion of CRC-S in the ensemble is redundant, although not detrimental.  However, when $\gamma$ is indeed sparse, CRC-L may perform poorly (this will be shown in the simulations of Section \ref{sec:simulation}; see Figure \ref{fig:crc}). Thus, while the characterization of CRC-L as a classifier trained on the latent variables $\bm{L}$ changes if $\gamma$ is not sparse, the accuracy of CRC does not.  For further comments, see Appendix \ref{recovering_L}.

\subsection{CRC-S: Training a classifier on the signal of primary interest}

We next build a classifier on $\bm{S}$.  DLDA is a natural choice for this classifier, since $S$ is normal with diagonal covariance.  Since neither $\bm{S}$ nor $S$ are known, the first step in training CRC-S is to obtain estimates $\hat{\bm{S}}$ and $\hat{S}.$ We propose two algorithms for doing so: residualization to obtain $\hat{S}$ and cross-residualization to obtain $\hat{\bm{S}}.$

\subsubsection{Residualization} \label{subsec:residualization}

Recovering $S$ for an out-of-sample observation is one of the key challenges of our approach. Even if $\bm{S}$ were known in the training data, a classifier trained on $\bm{S}$ would not be useful unless we could also recover $S$ for an out-of-sample observation $Z.$ 
%Both $\bm{S}$ and $S$ are needed in order to train and apply the meta-classifier we ultimately wish to fit.  
In this subsection, we focus on how to recover $S$.

Suppose that $L$ were observed (for both the training and target observations). Additionally, suppose $\gamma$ were known. In this case we could recover $S$ in a straight-forward manner. Since $\bm{Z}-\bm{T}\gamma = \bm{L}\alpha + \epsilon,$ an estimate of $\alpha$ could be obtained by regressing $\bm{Z}-\bm{T}\gamma$ onto $\bm{L}.$ Having obtained $\hat{\alpha}$, we could residualize the latent variables from $Z$ to obtain an estimate of $S$, %i.e.,
\[ \hat{S} = Z - L\hat{\alpha}. \]

However, since neither $L$ nor $\gamma$ are known this approach must be modified.  
With regard to $L$, one option would be to estimate $\bm{L}$ and $L$ directly by projecting $\bm{Z}$ and $Z$ onto the first several principal components of $\bm{Z}$.  We could then replace $\bm{L}$ and $L$ by their estimates in the procedure described above, and let $\hat{S} = Z - \hat{L}(\hat{\bm{L}}'\hat{\bm{L}})^{-1}\hat{\bm{L}}'(\bm{Z} - \bm{T}\gamma)$.  Since $\hat{\bm{L}}$ and $\hat{L}$ are obtained by projecting $\bm{Z}$ and $Z$ onto the principal components of $\bm{Z}$, this may be viewed as performing a principal components regression in which the training set predictors are $\bm{Z}$, the training set response variables are $\bm{Z}-\bm{T}\gamma$, and the resulting fitted model is applied to $Z$ in order to predict $L\alpha$; the predicted $L\alpha$ is then subtracted from $Z$ to give $\hat{S}$.  As discussed in the previous section, one challenge of PCR is selecting the number of principal components to use.  However, when $p \gg n$ good out-of-sample performance may be achieved by simply using all $n$ principal components.  As in the previous section, we adopt this approach both because it avoids the need to obtain an explicit estimate of $r$ and because it provides a simple closed form solution:
\begin{equation}
    \hat{S} = Z - Z\bm{Z}'(\bm{Z}\bm{Z}')^{-1}(\bm{Z} - \bm{T}\hat{\gamma})  \label{shatdef}
\end{equation}
where $\hat{\gamma}$ is an estimate of $\gamma$ (discussed below).  Another advantage of using all $n$ principal components is that it closely parallels the approach taken in CRC-L.  This is advantageous because our ultimate goal in computing $\hat{S}$ is to obtain any residual predictive information in $Z$ that is not already captured by CRC-L.  

With regard to estimating $\gamma$, several options are available.  Importantly, any estimate of $\gamma$ must take into account the presence of the latent variables.  In particular, simply taking the difference in class means, i.e., letting $\hat{\gamma} = (\bm{T}'\bm{T})^{-1}\bm{T}'\bm{Z}$, is not a viable option because this estimate will be biased if $\bm{L}$ is correlated with $\bm{T}$.  Estimators that account for $\bm{L}$ can be found in the batch effects normalization literature.  The estimator we use
\[
    \hat{\gamma} = [\bm{T}'\inv{(\bm{ZZ}')}\bm{T}]^{-1}\bm{T}'\inv{(\bm{ZZ}')}\bm{Z}
\] 
is from \citet{gagnon2013removing}, and has the advantage of having a simple closed form expression.  This estimator may be seen as approximating the ordinary least squares estimate of $\gamma$ from a regression in which $\bm{L}$ is known (Appendix \ref{sec:gammahat}).  Other estimates of $\gamma$, similar in spirit, include those obtained via surrogate variables analysis (SVA) \citep{leek2007capturing, leek2008general}, the confounder adjusted testing and estimation framework (CATE) \citep{wang2017confounder}, as well as several others \citep{listgarten2010correction, gagnonbartsch2012using, sun2012multiple, gerard2021unifying}.

\subsubsection{Cross-residualization} \label{subsec:cross-residualization}

In addition to recovering $S$ for an out-of-sample observation, we also wish to recover $\bm{S}$ in the training data.  A natural approach would be simply to apply the residualization procedure to the training data as well.  A complication is that the residualization procedure uses PCR with all $n$ principal components, and although this works well for out-of-sample observations, it massively overfits the training data.  Indeed, substituting $\bm{Z}$ for $Z$ in (\ref{shatdef}) to obtain $\hat{\bm{S}}$ would yield $\hat{\bm{S}} = \bm{Z} - \bm{Z}\bm{Z}'(\bm{Z}\bm{Z}')^{-1}(\bm{Z} - \bm{T}\hat{\gamma}) = \bm{T}\hat{\gamma}$, a rank one matrix in which each column (predictor) is perfectly correlated with $\bm{T}$.  This may be interpreted as the PCR overfitting to the $\bm{\epsilon}$ term in the training data, and therefore residualizing out the $\bm{\epsilon}$ term along with $\bm{L}\alpha$.

However, it is necessary to preserve the $\bm{\epsilon}$ term in order to properly train a classifier.
For example, in the DLDA classifier that we fit below, it is necessary to preserve the $\bm{\epsilon}$ term in order to estimate the individual feature variances $\sigma^2_1, \ldots, \sigma^2_p$.
More generally, overfitting in the training data introduces an asymmetry between the predictors $\hat{\bm{S}}$ on which the classifier is trained and the predictors $\hat{S}$ on which the classifier is applied.  As a result of this asymmetry, we would not expect a classifier trained on $\hat{\bm{S}}$ to generalize well to $\hat{S}$.

We address this problem by applying the residualization procedure to the training data in a leave-one-out  manner.  For each $i \in \{1 \ldots n\}$ we let 
\begin{equation}
    \hat{S}_i = Z_i - Z_i\bm{Z}'_{-i}(\bm{Z}_{-i}\bm{Z}'_{-i})^{-1}(\bm{Z}_{-i} - \bm{T}_{-i}\hat{\gamma}^{(i)})  \label{sihatdef}
\end{equation}
where $\hat{\gamma}^{(i)} = [\bm{T}_{-i}'\inv{(\bm{Z}_{-i}\bm{Z}_{-i}')}\bm{T}_{-i}]^{-1}\bm{T}_{-i}'\inv{(\bm{Z}_{-i}\bm{Z}_{-i}')}\bm{Z}_{-i}$.  We then let $\hat{\bm{S}} = [\hat{S}_1', \ldots, \hat{S}_n']'$.  By computing $\hat{\bm{S}}$ in this leave-one-out manner, we put $\hat{\bm{S}}$ and $\hat{S}$ on equal footing; each row of $\hat{\bm{S}}$ is computed in a manner analogous to the manner in which $\hat{S}$ is computed.  As a result, we can expect a classifier trained on $\hat{\bm{S}}$ to generalize well to $\hat{S}$.  We refer to this leave-one-out approach as \emph{cross-residualization}.

\emph{Comment.}  An alternative strategy to address overfitting in the training data would be to explicitly estimate $r$ when performing the PCR in the residualization procedure.  If $\hat{r} \ll n$ this would reduce the overfitting.  However, we prefer the cross-residualization strategy for multiple reasons.  Firstly, as noted previously, estimating $r$ is a challenging problem.  Secondly, for any $\hat{r}$ we would expect at least some degree of overfitting to the training data, introducing asymmetry between the training and target data, thereby impacting the ability of the classifier to generalize to the target data.  Finally, we note that because $(\ref{sihatdef})$ has a simple closed form solution, and in particular because $(\bm{Z}_{-i}\bm{Z}'_{-i})^{-1}$ can be computed using a rank-one downdate, cross-residualization can be implemented in a computationally efficient manner.  Thus, cross-residualization provides an approach that does not require selecting a tuning parameter, is computationally efficient, and guarantees that the training predictors $\hat{\bm{S}}$ and target predictors $\hat{S}$ are on equal footing.

\subsubsection{DLDA}

Cross-residualization effectively removes the sources of variation common across all features, and the columns of $\hat{\bm{S}}$ are therefore approximately decorrelated.  Thus when fitting a classifier to $\hat{\bm{S}}$ we may take advantage of this decorrelation and fit by DLDA.  Feature selection is an important aspect of fitting the DLDA classifier.  Because cross-residualization effectively removes the sources of variation common across all features, we wish to select features based on their individual predictive effects beyond the common latent effect. Selecting too many null features may render the ensemble step less useful. Since the columns of $\hat{\bm{S}}$ are approximately decorrelated, we utilize a simple marginal screening scheme in which only the features with the $N$ smallest $p$-values are used for classification (i.e., the `top' $N$ features). Here, $N \in \{ 1, \ldots, p \}$ is a tuning parameter that is selected via a grid search. Details are in Appendix \ref{feature_selection}.

\subsection{Fitting the ensemble}  \label{sec:ensemble}

There are many ways in which we might combine CRC-S and CRC-L into an ensemble.  Voting is a popular approach \citep{yang2010review}, but here there are only two classifiers to be combined. Instead, we adopt elements of the super learning approach of \citet{polley2011super}. In the super learner, the training data are split into several folds, and each classifier in the ensemble is fit to each fold. 
The resulting fits are used to estimate the optimal weighted combination of classifiers. These estimated weights minimize cross-validated error over all possible weighted linear combinations of the classifiers. The ensemble classifier is then obtained by fitting each of the classifiers to the entire training set and combining them with the estimated weights. 

The CRC is well-suited to the super learning approach, since cross-residualization is already a leave-one-out procedure. In fact, it is imperative to fit the CRC ensemble using some type of cross-validated approach, otherwise we may overfit the training data. To fit the CRC ensemble, we score the training observations using CRC-S and CRC-L in a leave-one-out manner and then train a meta-classifier on these scores. Under our model, these scores are Gaussian conditional on the class label and we therefore use LDA for the meta-classifier.  Assembling all steps together, we arrive at Algorithm \ref{alg:CRC}, which summarizes the CRC.

\begin{figure}[htp]
\begin{algorithm}[H]
  \SetAlgoLined\DontPrintSemicolon
  \KwData{$\bm{Z}_{n \times p}$, $\bm{T}_{n \times 1}$, $Z_{1 \times p}$}
  \KwResult{$\hat{T} \in \{\pm 1\}$}
  \def\NoNumber#1{{\def\alglinenumber##1{}\State #1}\addtocounter{ALG@line}{-1}}

  \SetKwProg{myalg}{Residualization}{}{}
  \myalg{}{
  \KwData{$\bm{Z}_{n \times p}$, $\bm{T}_{n \times 1}$, $Z_{1 \times p}$}
  \KwResult{$\hat{S}_{1 \times p}$}
   $\hat{\gamma} \leftarrow [\bm{T}'\inv{(\bm{ZZ}')}\bm{T}]^{-1}\bm{T}'\inv{(\bm{ZZ}')}\bm{Z}$\;
   $\widehat{L\alpha} \leftarrow Z (\bm{ZZ}')^{-1} (\bm{Z}-\bm{T}\hat{\gamma})$\;
   \KwRet $\hat{S} \leftarrow Z - \widehat{L\alpha}$\;}{}

\SetKwProg{myalg}{Cross-residualization}{}{}
  \myalg{}{
  \KwData{$\bm{Z}_{n \times p}$, $\bm{T}_{n \times 1}$}
  \KwResult{$\hat{\bm{S}}_{n \times p}$}
   \For{$i$ in 1, \ldots, n}{
  	$\hat{S}_i \leftarrow$ result of residualization algorithm applied to $(\drop{\bm{Z}}, \drop{\bm{T}}, Z_i)$\;}
   \KwRet $\hat{\bm{S}} \leftarrow [\hat{S}_1', \ldots, \hat{S}_n']'$ \;}

\SetKwProg{myalg}{CRC-L}{}{}
  \myalg{}{
  \KwData{$\bm{Z}_{n \times p}$, $\bm{T}_{n \times 1}$, $Z_{1 \times p}$}
  \KwResult{$\scl \in \mathbb{R}$}
  	Use PC-LDA (with $n$ principal components) to train a classifier on  $(\bm{Z}, \bm{T}),$ resulting in weight vector $\wl$\;
  	$\scl \leftarrow Z\wl$\;
   \KwRet $\scl$ \;}

\SetKwProg{myalg}{CRC-S}{}{}
  \myalg{}{
  \KwData{$\hat{\bm{S}}_{n \times p}$, $\bm{T}_{n \times 1}$, $\hat{S}_{1 \times p}$}
  \KwResult{$\scs \in \mathbb{R}$}
  Use DLDA to train a classifier on  $(\hat{\bm{S}}, \bm{T}),$ resulting in weight vector $\ws$\;
  $\scs \leftarrow \hat{S}\ws$\;
   \KwRet $\scs$ \;}

\SetKwProg{myalg}{Ensemble classifier}{}{}
  \myalg{}{
  \KwData{ $\bm{Z}_{n \times p}$, $\hat{\bm{S}}_{n \times p}$, $\bm{T}_{n \times p}$, $\scl$, $\scs$}
  \KwResult{$\hat{T} \in \{\pm 1\}$}
  \For{$i$ in 1, \ldots, n}{
  	$\scli \leftarrow$ result of CRC-L applied to $(\drop{\bm{Z}}, \drop{\bm{T}}, Z_i)$\;
  	$\scsi \leftarrow $ result of CRC-S applied to $(\drop{\hat{\bm{S}}}, \drop{\bm{T}}, \hat{S}_i)$\;}
  Use LDA to train a classifier $\hat{c}$ on predictors
  $\{ (\scli, \scsi) \}_{i=1}^n, $ and response $\bm{T}$\;
  \KwRet $\hat{c}(\scl, \scs)$\;}
  \caption{Cross-residualization classifier (CRC)}
  \label{alg:CRC}
\end{algorithm} 
\begin{justify}
\emph{Note: Some details omitted for clarity.  In particular, CRC-S has a feature selection step, and the implementations of both CRC-L and CRC-S differ slightly when applied to $n-1$ observations rather than $n$.  See Appendices \ref{crcl_implementation} and \ref{feature_selection}.}
\end{justify} 
\end{figure}

An alternative interpretation of the CRC arises if we unpack Algorithm \ref{alg:CRC} and examine how the CRC acts on $Z$ instead of the discriminant scores $\scs$ and $\scl$. Note that $\scs$ and $\scl$ arise as linear functions of $Z.$ For $\scl$, this is clear. For $\scs$, observe that (\ref{shatdef}) may be rewritten as $\hat{S} = Z [I - \bm{Z}'\inv{(\bm{Z}\bm{Z}')}(\bm{Z}-\bm{T}\hat{\gamma})]$ and hence
\begin{align*}
  \scs
  &= \hat{S}\ws\\
  &= Z [I - \bm{Z}'\inv{(\bm{Z}\bm{Z}')}(\bm{Z}-\bm{T}\hat{\gamma})] \ws
\end{align*}
is a linear function of $Z$.  In addition, $\hat{c}$ is itself a linear classifier and can be expressed as $\hat{c}(\scl, \scs) = \sign{b_1 \scs + b_2 \scl}$ where $b_1$ and $b_2$ are scalar weights. Therefore \begin{align*}
\hat{c}(s_S, s_L) &= \mathrm{sign}\left\{b_1 Z\left[I - \bm{Z}'\inv{(\bm{Z}\bm{Z}')}(\bm{Z}-\bm{T}\hat{\gamma})\right] \ws + b_2 Z \wl\right\}\\
  &= \sign{Z\wc}
\end{align*}
where $\wc = b_1 [I - \bm{Z}'\inv{(\bm{Z}\bm{Z}')}(\bm{Z}-\bm{T}\hat{\gamma})] \ws + b_2 \wl$. The expression for $\wc$ shows that the CRC is itself a linear classifier that weights each feature of an observation by a weighted average of the CRC-S weights (meant to capture the sparse signals) and CRC-L weights (meant to capture the dense, low-rank latent signals), where the relative weighting is determined by the relative predictive value of the individual classifiers. We therefore see that features may receive large weights due to the discriminative information they carry about the sparse signals, the latent signals, or both.

%%%%%%%%%%%%%%%%%%%%%%%%%%%%%%%%%%%%%%%%%%%%%%
%%%% Section 4
%%%%%%%%%%%%%%%%%%%%%%%%%%%%%%%%%%%%%%%%%%%%%%

\section{Simulations}
\label{sec:simulation}

We perform several simulations to illustrate the inner workings of the CRC and to compare the CRC to other classifiers which are frequently used in genomic applications.

We generate data according to the simple, uncorrelated, and correlated models from Section \ref{sec:model-motiv}. For all three models, we let $\Sigma = I_p$ and $\gamma = (\frac{1}{\sqrt{3}}, \frac{1}{\sqrt{3}}, \frac{1}{\sqrt{3}}, 0, \ldots, 0)$. For the uncorrelated and correlated models, we set $k=3$ and $\Psi = I_{k \times k}$, and generate i.i.d.\ $\alpha_{ij}$ from a standard normal distribution. In the correlated model, we additionally set $\eta = (\frac{1}{\sqrt{3}}, \frac{1}{\sqrt{3}}, \frac{1}{\sqrt{3}})$. Note that the Bayes optimal accuracy rate is $\Phi(\sqrt{\gamma'\gamma})$ for a classifier trained on $S$ and $\Phi(\sqrt{\eta'\eta})$ for a classifier trained on $L.$ For a classifier trained on $(S,L),$ the Bayes optimal accuracy rate is $\Phi(\sqrt{\gamma'\gamma + \eta'\eta}).$ Thus with our parameter choices, the Bayes optimal accuracy rate for a classifier trained on $(S,L)$ is $\Phi(1)$ under the simple model, $\Phi(1)$ under the uncorrelated model, and $\Phi(\sqrt{2})$ under the correlated model.

\begin{figure}[htb]
    \centering
    \includegraphics[width=.8\linewidth]{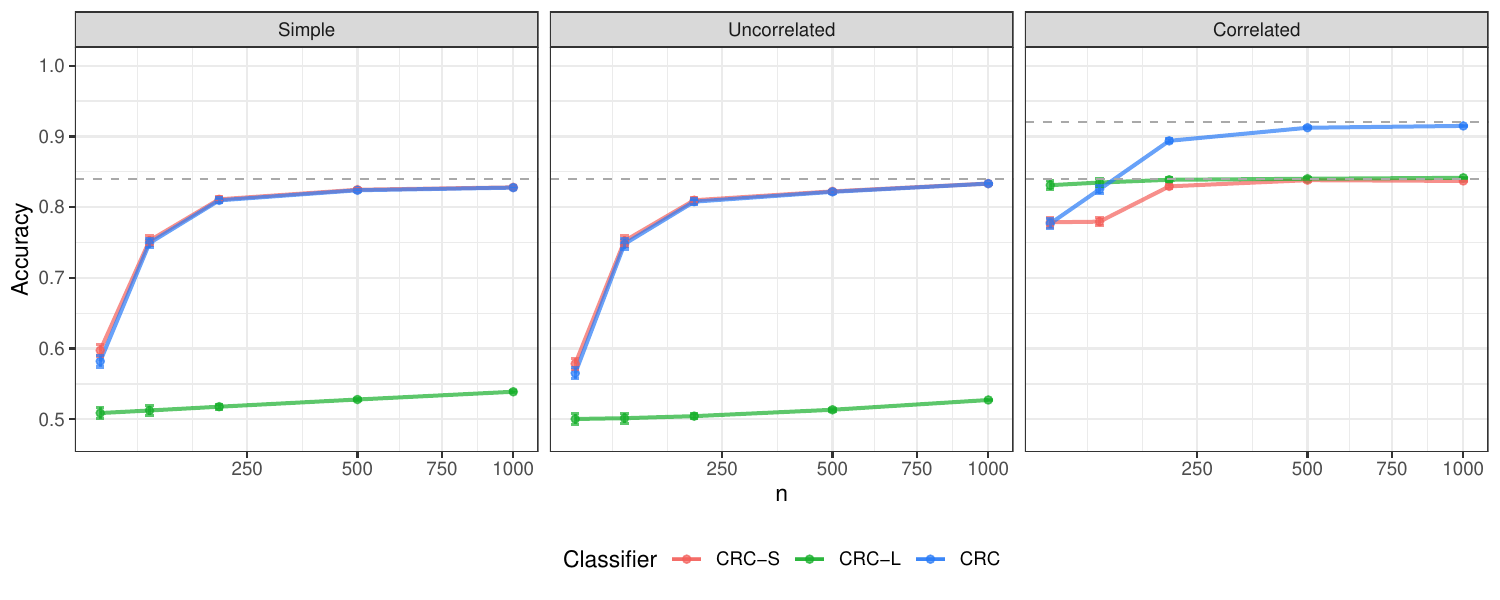}
    \caption{Mean accuracies of CRC, CRC-S, and CRC-L  for $p=\bignumber{100000}$. Sample size $n$ is depicted on a square-root scale. The dashed horizontal lines are at $\Phi(1)$ and $\Phi(\sqrt{2})$ and indicate Bayes optimal accuracy rates (as detailed in the text).}
    \label{fig:crc}
\end{figure}

For each model, we generate a balanced class label vector of dimension $n$ and a feature matrix of dimension $n \times p,$ for varying $n$ and $p$. We let $n$ range from 50 to 1000 and consider $p=\bignumber{20000}, \bignumber{100000},$ and $\bignumber{500000},$ roughly corresponding to the number of features found in various types of ``omics'' datasets. For each $n$ and $p,$ we replicate this procedure several times and average accuracy rates for CRC, CRC-S, and CRC-L across replications. For computational feasibility, we vary the number of replications with $n$ (see Section \ref{sim_supplement} of the supplement). Results for $p=\bignumber{100000}$ are depicted in Figure \ref{fig:crc}; results for for $p=\bignumber{20000}$ and $p=\bignumber{500000}$ can be found in Section \ref{sim_supplement} of the supplement.

To see how the CRC ensemble works to improve classification accuracy, we first look to the performance of its component classifiers, CRC-L and CRC-S. CRC-L has an accuracy rate close to 50\% under the simple and uncorrelated models, which can be attributed to the sparsity of $\gamma$ and the additional noise contributed by the latent variables in the uncorrelated case. The accuracy of CRC-L does increases slightly as $n$ increases, as is expected. Although CRC-L performs poorly under the simple and uncorrelated models, it has accuracy close to the Bayes optimal rate (for $L$) under the correlated model, even when $n$ is relatively small. Looking to the other component classifier, we see that CRC-S performs well across all models. For large $n,$ CRC-S has accuracy close to the Bayes optimal rate for $S.$ In particular, CRC-S exhibits similar performance under the simple and uncorrelated models, suggesting that cross-residualization is providing a reasonable estimate of $\bm{S},$ allowing CRC-S to effectively pick up on the sparse signal. 

CRC behaves as expected in the simple and uncorrelated cases, with accuracies on par with CRC-S. However, it is the correlated case that is particularly illustrative. For large $n,$ CRC-S and CRC-L have accuracies close to the Bayes accuracy rates for $S$ and $L,$ respectively, but both of these accuracies are less than the Bayes accuracy rate for $(S,L).$ The accuracy of the CRC ensemble, however, is close to the Bayes accuracy rate for $(S,L)$ for large $n,$ suggesting that the CRC is making more efficient use of the signal in the data by considering the sparse signals and dense latent signals separately. Whereas the uncorrelated case demonstrates that improved recovery of the sparse signals can improve classification accuracy, the correlated case demonstrates that there are potential information gains to be made beyond better sparse recovery when dense latent signals exist.

\begin{figure}[htb]
    \centering
	\includegraphics[width=.8\linewidth]{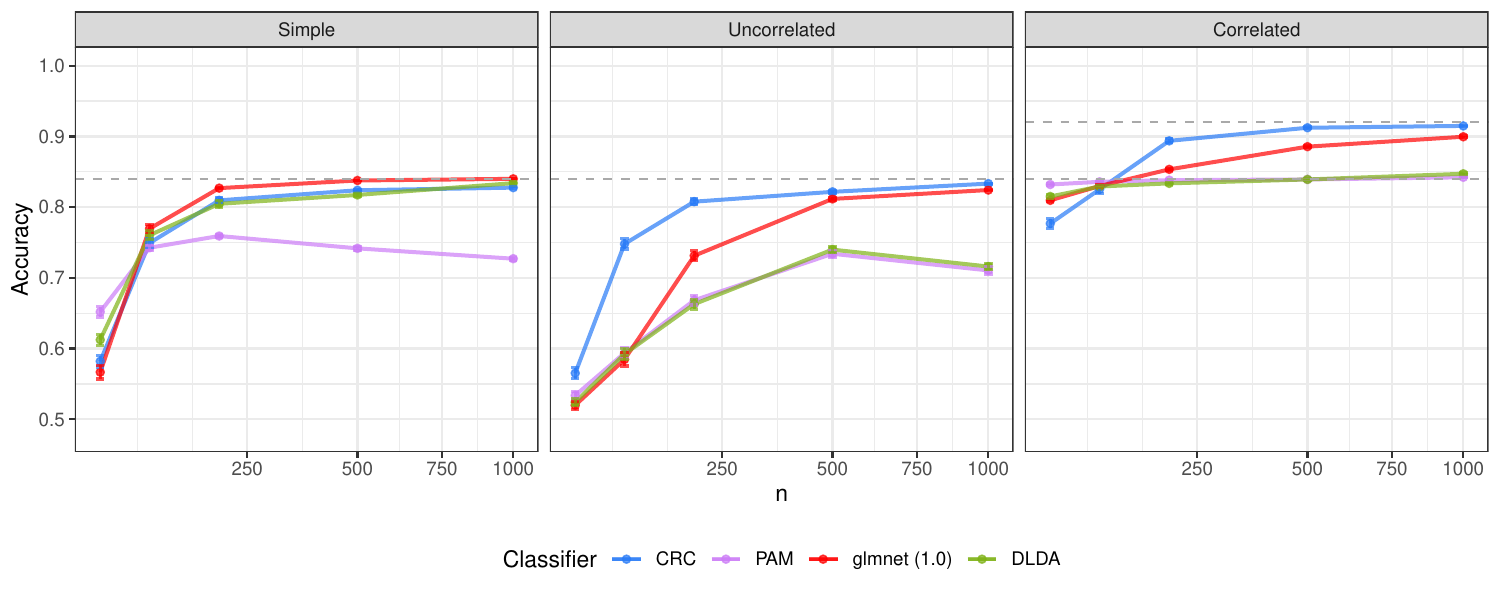}
    \caption{Mean accuracies of glmnet, PAM, and CRC $(p=\bignumber{100000})$. Sample size $n$ is depicted on a square-root scale. The dashed horizontal lines are at $\Phi(1)$ and $\Phi(\sqrt{2})$ and indicate Bayes optimal accuracy rates (as detailed in the text).}
    \label{fig:all_clfs}
\end{figure}

In Figure \ref{fig:all_clfs}, we repeat the simulation with other classifiers commonly used in genomics: penalized logistic regression (glmnet) \citep{zou2005elasticnet}, DLDA \citep{dudoit2003classification}, and nearest shrunken centroids (PAM) \citep{tibshirani2002diagnosis}. Both glmnet and PAM have shrinkage parameters which are tuned via cross-validation. The glmnet classifier has an additional parameter, the elastic net mixing parameter $\alpha$. We set $\alpha$ to various values from 0 to 1 since simulation results are potentially sensitive to the value of this parameter. For visual simplicity, we include only the results for $\alpha=1$ in Figure \ref{fig:all_clfs}; this is the value of $\alpha$ for which glmnet generally performs best. Results for glmnet at other values of $\alpha$ can be found in in Section \ref{sim_supplement} of the supplement. We use our own implementation of DLDA, which includes a feature selection step that is identical to the feature selection step in CRC-S. As a result, DLDA and CRC-S differ only in that DLDA is fit to $\bm{Z}$ whereas CRC-S is fit to $\hat{\bm{S}}.$

Figure \ref{fig:all_clfs} displays the results. Under the simple model, the classifiers perform similarly, with the exception of PAM, for which accuracy appears to degrade as the sample size increases. We believe that this is related to PAM selecting a large number of features in the simple setting (see Table \ref{tab:Nsel}). Under the uncorrelated and correlated models, however, the CRC appears to offer substantial gains over the other methods, particularly for moderate sample sizes ($n=100$, $200$). PAM and DLDA perform similarly under the correlated and uncorrelated models, which is perhaps expected since PAM is essentially DLDA with an $\ell_1$ penalty to perform feature selection.

It is interesting to compare CRC-S and DLDA under the three models. Recall that in our simulation, the two methods differ only in that DLDA is fit to $\bm{Z}$, whereas CRC-S is fit to $\hat{\bm{S}}.$ As expected, DLDA behaves much like CRC-S under the simple model. However, this is no longer the case when latent factors are present. Under the uncorrelated model, DLDA is outperformed by CRC-S despite both classifiers selecting similarly small number of features (see Table \ref{tab:Nsel}), suggesting that feature selection alone is not necessarily effective when the optimal feature weights are nonsparse. As noted in Section \ref{sec:model-motiv}, sparsity in $S$ does not necessarily imply sparsity in $Z,$ even if $L$ and $T$ are uncorrelated. Under the correlated model, DLDA performs similarly to CRC-S, but both are outperformed by the CRC ensemble, highlighting the effectiveness of cross-residualization when used in conjunction with an ensemble strategy.

\begin{table}
\caption{Median number of features selected ($p=\bignumber{100000}).$ `S', `U', and `C' denote the Simple, Uncorrelated, and Correlated settings.}
\label{tab:Nsel}
\centering
\vspace{0.2cm}
\begin{tabular}{lrrrrrrrrrrrr}
\toprule
\multicolumn{1}{c}{ } & \multicolumn{3}{c}{glmnet ($\alpha = 1$)} & \multicolumn{3}{c}{PAM} & \multicolumn{3}{c}{DLDA} & \multicolumn{3}{c}{CRC-S} \\
\cmidrule(l{3pt}r{3pt}){2-4} \cmidrule(l{3pt}r{3pt}){5-7} \cmidrule(l{3pt}r{3pt}){8-10} \cmidrule(l{3pt}r{3pt}){11-13}
  & S & U & C & S & U & C & S & U & C & S & U & C\\
\midrule
$n=50$ & 0.0 & 0 & 10.0 & 32.5 & 9.0 & 8.0 & 9.5 & 3 & 11.0 & 11 & 11 & 23\\
$n=100$ & 7.0 & 1 & 16.5 & 145.0 & 11.0 & 14.0 & 4.0 & 2 & 6.0 & 4 & 4 & 8\\
$n=200$ & 7.5 & 29 & 27.5 & 469.5 & 10.0 & 35.0 & 4.0 & 3 & 3.0 & 4 & 4 & 4\\
$n=500$ & 5.0 & 72 & 49.0 & 1915.0 & 3.5 & 39.0 & 6.0 & 3 & 3.5 & 6 & 6 & 4\\
$n=1000$ & 3.0 & 94 & 86.0 & 4954.0 & 4.5 & 443.5 & 4.0 & 3 & 7.0 & 6 & 4 & 4\\
\bottomrule
\end{tabular}
\end{table}

%%%%%%%%%%%%%%%%%%%%%%%%%%%%%%%%%%%%%%%%%%%%%%
%%%% Section 5
%%%%%%%%%%%%%%%%%%%%%%%%%%%%%%%%%%%%%%%%%%%%%%

\section{Applications to Genomic Data}
\label{sec:real-data}

We apply the CRC to a variety of publicly available genomic datasets.  These datasets cover a broad range of phenotypes, including Alzheimer's disease, asthma, cancer, Crohn's disease, fetal alcohol spectrum disorder (FASD), and sepsis \citep{gasparoni2018dna, nicodemus2016dna, xu2013identification, haberman2019age, somineni2019blood,cobben2018dna, washburn2019t}. The datasets also encompass several technologies and include gene expression data from multiple platforms (Illumina HiSeq 2000 and 2500, and Affymetrix HG U133 Plus 2.0 microarray) as well as methylation data from two platforms (Illumina Infinium 450K and EPIC BeadChip arrays).  See Table \ref{table:datasets} for a summary, and Section \ref{supp:datasets} of the supplement for additional details. We select these particular datasets because they have a relatively large number of samples within each class and a clearly defined phenotype.

\begin{table}%[!p]
\caption{Datasets. In the ``Classes'' column, the number of samples in each class is given in parentheses. The Alzheimer's and Sepsis datasets contain multiple samples per subject (individual person), so we additionally give the number of subjects. See Section \ref{supp:datasets} of the supplement for details.
\label{table:datasets}}
\centering
\vspace{0.2cm}
\scalebox{0.9}{
\begin{tabular}{lllrrll}
\toprule
Name & Type & Platform & $n$ & $p$ & Classes\\
\midrule
Alzheimer's         & Methyl.   & HM 450k       & 190    & 485,512   & Disease (106; 54), Control (84; 42) \\
Asthma              & Methyl.   & HM 450k       & 115   & 485,512   & Asthma (74), Control (41) \\
Colorectal Cancer   & Expr.     & U133 Plus 2.0 & 200   & 54,675    & Cancer (100), Normal (100)\\
Crohn's (Methyl.)   & Methyl.   & EPIC          & 402   & 504,790   & CD (328), Non-IBD Control (74) \\
Crohn's (Expr.)     & Expr.     & HiSeq 2000    & 304   & 13,151    & CD (254), Non-IBD (50) \\
FASD                & Methyl.   & HM 450k       & 103   & 485,512   & FASD (39),  Control (64) \\
Sepsis              & Expr.     & HiSeq 2500    & 217   & 27,670    & Healthy (58; 20), \\
&&&&& Critically ill non-septic (63; 22),\\
&&&&& Sepsis (96; 29)\\
\bottomrule                 
\end{tabular}
}
\end{table}

The percentage of variance explained by the top ten principal components is given in Table \ref{table:pct_var_explained}. Scree plots and additional details can be found in Section \ref{supp:datasets} of the supplement. For each dataset, the top ten principal components capture a large percentage of the variation in the data, which we interpret as evidence of a common effect induced by latent variables.

\begin{table}[b]
\caption{Percentage of variance explained by the first 10 principal components.
\label{table:pct_var_explained}}
\centering
\vspace{0.2cm}
\begin{tabular}{lr}
\toprule
Dataset & \% of variance explained \\
\midrule
Alzheimer's & 75.1\% \\
Asthma & 54.6\% \\
Crohn's (Expression) & 75.6\% \\
Crohn's (Methylation) & 50.2\% \\
Colorectal Cancer & 78.5\% \\
FASD & 53.7\% \\
Sepsis & 62.4\% \\
\bottomrule                 
\end{tabular}
\end{table}

We compare the CRC to the same classifiers from Section \ref{sec:simulation}.
For each dataset, we assess the performance of each classifier as follows. Let $n_0 = \min\{n_1, n_2\}$ where $n_1$, $n_2$ are the class sizes (number of observations in each class). To form the training set, we randomly sample $\floor{0.8 \cdot n_0}$ observations from each class. Of the remaining observations, we sample $n_0 - \floor{0.8 \cdot n_0}$ from each class to form the test set. This results in balanced training and test sets, so that the baseline accuracy rate is 50\% across datasets and interpretation of accuracies is not complicated by class imbalances.  
We create 200 train-test splits in this way and compute the mean test accuracy of each classifier across replications (Table \ref{tab:results}).
For the Alzheimer's and Sepsis datasets, we modify the procedure slightly to account for the fact that there are multiple observations per individual; see Section \ref{supp:additional_analysis} of the supplement.

The CRC generally performs well relative to other classifiers. We see that CRC-L generally performs well across the datasets, suggesting that there may be dense latent factors that are predictive of class label. CRC-S generally outperforms DLDA, highlighting the value of (cross-) residualization, and suggesting once again that the signal of primary interest may be sparse in $S$ but dense in $Z$. As with the simulations of Section \ref{sec:simulation}, the relative performance of CRC-S and DLDA on these datasets suggests that feature selection in conjunction with cross-residualization can be more effective than feature selection alone when latent variation is present. Overall, the CRC generally performs as well as the better of its two component classifiers. 

\begin{table}%[!p]
\caption{Results on various genomic datasets. Standard errors are provided in Section \ref{supp:additional_analysis} of the supplement. Boldface denotes highest accuracy rate.
\label{tab:results}}
\centering
\vspace{0.2cm}
\begin{tabular}{lrrrr|rr}
\toprule
  & glmnet & PAM & DLDA & CRC & CRC-S & CRC-L\\
\midrule
Alzheimer's & 0.74 & 0.71 & 0.70 & \textbf{0.78} & 0.78 & 0.80\\
Asthma & 0.77 & \textbf{0.79} & 0.75 & 0.77 & 0.76 & 0.77\\
Colorectal Cancer & 0.91 & 0.74 & 0.75 & \textbf{0.94} & 0.94 & 0.88\\
Crohn's (Expr.) & 0.90 & 0.87 & 0.88 & \textbf{0.92} & 0.89 & 0.92\\
Crohn's (Methyl.) & 0.87 & 0.86 & 0.85 & \textbf{0.88} & 0.88 & 0.79\\
FASD & 0.76 & 0.79 & 0.76 & \textbf{0.83} & 0.83 & 0.79\\
Sepsis (Healthy vs. Sepsis) & \textbf{0.95} & 0.90 & 0.91 & \textbf{0.95} & 0.95 & 0.94\\
Sepsis (Crit. Ill. vs. Sepsis) & 0.70 & 0.70 & 0.69 & \textbf{0.78} & 0.76 & 0.77\\
Sepsis (Crit. Ill. vs. Healthy) & 0.85 & 0.80 & 0.80 & \textbf{0.89} & 0.89 & 0.83\\
\bottomrule
\end{tabular}
\end{table}

Another promising result is that the CRC appears to perform well within subgroups of interest. In the Alzheimer's dataset, there are four types of samples: purified glia cells, purified neuron cells, bulk samples taken from the temporal cortex, and bulk samples taken from the frontal cortex. In Section \ref{supp:additional_analysis} of the supplement, we report accuracy rates within each subgroup for the analysis performed in this section. Notably, the CRC performs as well or better than the other classifiers within each cell type. For certain cell types and classification tasks, there is a marked improvement in accuracy rates. We see similar results in the Sepsis dataset, in which samples occur across three cell types (CD4, CD14, and CD8).

%%%%%%%%%%%%%%%%%%%%%%%%%%%%%%%%%%%%%%%%%%%%%%
%%%% Section 6
%%%%%%%%%%%%%%%%%%%%%%%%%%%%%%%%%%%%%%%%%%%%%%

\section{Discussion}
\label{sec:discussion}

The CRC ensemble derives its strength from the separation of the sparse signal of interest and dense latent signals. That is, while ensembles in general typically offer some benefit by combining multiple classifiers, the CRC fully exploits this benefit by combining classifiers trained on different sources of information. In particular, we are able to better leverage information contained in the sparse signal, which may be obscured by the dense latent signals.

In addition to its statistical performance, the CRC offers several practical advantages, many of which stem from its modularity. There are four individual prediction algorithms used within the CRC --- the PC-LDA in CRC-L; the PCR in the residualization step; the DLDA in CRC-S; and the LDA that defines the ensemble. There are several distinct benefits to this modular structure.  In particular, modularity makes it easier to conceptualize an approach as well as easier to implement and debug \citep{gerard2021unifying}.  In addition, modularity produces additional opportunities for diagnostics, as there are intermediate outputs that can be inspected and visualized.  Finally, because individual algorithms are easily swapped out, the CRC can be readily refined and adapted to new settings.

For example, we may wish to modify CRC-L and replace PC-LDA with some other algorithm if there is reason to believe that the latent variables are not normally distributed. This might occur, for example, if there are subgroups within a class (e.g., subtypes of a cancer).  In this case, it may be desirable to fit CRC-L using PC-kNN or some other non-parametric method. To maintain the complementary nature of CRC-S and CRC-L, we might choose to use PC-kNN within the residualization and cross-residualization algorithms as well, so that the same signals captured by CRC-L are those residualized out of $Z$.

Improvements and adaptations of CRC-S are also possible. For example, we could do away with a marginal screening procedure for feature selection, and instead utilize shrinkage penalties.  We believe that such alternatives have the potential to improve the accuracy of CRC-S; the fact that glmnet outperforms CRC-S in the simple simulation (left panel of Figure \ref{fig:all_clfs}) suggests that there is indeed room for improvement.  In addition, CRC-S could be generalized to accommodate more general error structures.  
For example, even after removing the latent variables, correlations may remain between a few individual genes whose functions are tightly related.  In such cases, it may make sense to model $\Sigma$ as a sparse but not strictly diagonal matrix.

Finally, neither residualization nor cross-residualization require $\bm{T}$ to be a binary variable. By taking $\bm{T}$ to be continuous, we can extend CRC to the regression setting, provided that we modify CRC-S and CRC-L accordingly (i.e., substituting regression-based equivalents). Again, the modularity of the CRC makes this relatively straightforward from both a conceptual and practical standpoint.

The specific algorithms that we have chosen to use in the CRC also offer practical advantages of their own.  In particular, our implementations of PC-LDA and PCR use all $n$ principal components and therefore do not require the selection of a tuning parameter.  
The resulting methods are computationally efficient and have closed form solutions.  
In using all $n$ principal components, we take advantage of the fact that PCR and PC-LDA continue to have good out-of-sample predictions despite severely overfitting to the training data.
This strategy is made possible by the leave-one-out manner in which we apply these algorithms, which allows us to avoid issues that would otherwise arise due to the overfitting of the training data. 
In particular, the leave-one-out nature of cross-residualization allows us to preserve the $\bm{\epsilon}$ term, which is critical to properly fitting CRC-S.  
In addition, the leave-one-out manner in which we obtain the CRC-L and CRC-S scores allows us to properly weight CRC-L and CRC-S in the ensemble.
The net effect of this approach is that we replace tuning with over-parameterization and leave-one-out fits; we are hopeful that this general strategy may be useful in other contexts as well.
The leave-one-out fits within the ensemble provide another benefit; namely, built-in estimates of error rates for the individual components and the CRC as a whole.

Regardless of the specific implementation, we believe the analyses in this paper highlight the importance of accounting for the latent variables that are prevalent in genomic data. By doing so, weaker biological signals which may be less prominent but equally as important can be better incorporated into analyses of such data.

\section{Software and code}
\label{sec: software-code}
An R package implementation of the CRC, as well as code for reproducing the results in this paper, is publicly available at \url{https://github.com/yujiap/crc_code}.

\begin{appendix}
\section{Recovering the latent variables}\label{recovering_L} 
Let $\bm{Z}_{n \times p} = \bm{T}_{n \times 1}\gamma + \bm{L}_{n \times r}\alpha + \bm{\epsilon}_{n \times p}$  denote the training data.  This notation is introduced at the end of Section \ref{sec2discussion}.  Each row $i$ of $\bm{Z}$ is independently generated from the correlated latent factor model of Section \ref{correlatedmodel}, i.e.,
\[
	Z_i = T_i\gamma + L_i\alpha + \epsilon_i
\]
where $T_i \in \{\pm 1\}$ with $P(T=1) = \pi$ and $L_i \mid T_i \sim N(T_i\eta, \Psi)$, and where $\gamma_{1 \times p}$, $\alpha_{r \times p}$, $\eta_{1\times r}$ and $\Psi_{r \times r}$ are fixed (not random) parameters.  For simplicity, we assume here that $\epsilon_i \sim N(0, \sigma^2 I)$, i.e., all features have the same error variance $\sigma^2$.

Consider the limit where $n$ is fixed but $p \to \infty$.  The parameters $\gamma_{1 \times p}$ and $\alpha_{r \times p}$ must grow with $p$.  We assume that as they do so,
\begin{align}
    \gamma \gamma'/p &\to 0 \label{sparsegamma}\\
    \alpha \alpha'/p &\to \Lambda_{r \times r} \label{densealpha}
\end{align}
where $\Lambda_{r \times r}$ is a positive definite matrix.  Condition (\ref{sparsegamma}) reflects the sparsity of $\gamma$ and condition (\ref{densealpha}) reflects the density of $\alpha$.  It follows that 
$\bm{T}\gamma\gamma'\bm{T}'/p \overset{p}{\to} 0$,
$\bm{T}\gamma\alpha'\bm{L}'/p \overset{p}{\to} 0$,
$\bm{T}\gamma\epsilon'/p \overset{p}{\to} 0$,
$\bm{L}\alpha\alpha'\bm{L}'/p \overset{p}{\to} \bm{L}\Lambda\bm{L}'$,
$\bm{L}\alpha'\epsilon/p \overset{p}{\to} 0$, and
$\epsilon\epsilon'/p \overset{p}{\to} \sigma^2I$, and therefore that
\begin{equation}
    \bm{ZZ}'/p \overset{p}{\to} \bm{L}\Lambda\bm{L}' + \sigma^2I.
\end{equation}
Suppose for simplicity that the first $r$ eigenvalues of $\bm{L}\Lambda\bm{L}' + \sigma^2I$ are distinct.  Then by continuity, each of the first $r$ eigenvectors of $\bm{ZZ}'/p$ converges in probability to the respective eigenvector of $\bm{L}\Lambda\bm{L}' + \sigma^2I$ (see, e.g., \cite{ortega1990numerical}).  Note that the first $r$ eigenvectors of $\bm{L}\Lambda\bm{L}' + \sigma^2I$ are the same as the first $r$ eigenvectors of $\bm{L}\Lambda\bm{L}'$, and in particular that these eigenvectors are simply a linear transformation of $\bm{L}$.  That is, we may write $\tilde{\bm{L}} = \bm{L}Q$ where $\tilde{\bm{L}}$ denotes the first $r$ eigenvectors of $\bm{L}\Lambda\bm{L}' + \sigma^2I$ and $Q_{r \times r}$ is some invertible matrix.  Note also that the first $r$ left singular vectors of $\bm{Z}$ are the same as the eigenvectors of $\bm{ZZ}'/p$.  It therefore follows that if we  denote the first $r$ left singular vectors of $\bm{Z}$ by $\hat{\bm{L}}$, we have that $\hat{\bm{L}} \overset{p}{\to} \tilde{\bm{L}}$.

The matrix $Q$ is unidentifiable and may be viewed as a reparameterization of $\bm{L}$ and $\alpha$.  In particular, if we define $\tilde{\alpha} = Q^{-1}\alpha$, then $\tilde{\bm{L}}\tilde{\alpha} = \bm{L}\alpha$.  Thus, for the purposes of regressing out the $\bm{L}\alpha$ term as described in Sections \ref{sec2discussion} and \ref{subsec:residualization}, $\tilde{\bm{L}}$ is equivalent to $\bm{L}$.  Similarly, linear discriminant analysis (LDA) is invariant to linear transformation of the predictors, and thus for the purposes of PC-LDA described in Section \ref{subsec:CRC-L},  $\tilde{\bm{L}}$ is equivalent to $\bm{L}$. 

\emph{Comments:} Note that condition (\ref{sparsegamma}) does not correspond directly to a notion of sparsity.  In particular, $\gamma$ could be sparse, but with a few very large entries, in which case condition (\ref{sparsegamma}) might not hold.  However, if the entries of $\gamma$ are bounded, as we might expect in any realistic scenario, then this is not a concern.  More specifically, if the entries of $\gamma$ are bounded and if we let $p^+$ denote the number of non-null entries of $\gamma$, then $p^+/p \to 0$ implies condition (\ref{sparsegamma}).

Another scenario of interest is that $p^+/p \to \delta > 0$ and $\gamma \gamma'/p \to \zeta > 0$.  In this case, $\bm{T}$ could be recovered as well as $\bm{L}$ by the left singular vectors of $\bm{Z}$.  More specifically, if we let $\bm{W}_{n \times r + 1} = \begin{pmatrix} \bm{L} & & \bm{T} \end{pmatrix}$, then by an argument analogous to the one above, we could recover $\bm{W}$ perfectly in the limit, up to a linear transformation.  In this case, the PC-LDA classifier mentioned in Section \ref{subsec:CRC-L} would have perfect accuracy, since $\bm{T}$ itself is now one of the predictors.  We emphasize that this applies only in the limit $p \to \infty$.  (However, $n$ could be any finite value, as long as $n > r+1$.)  We also note that this assumes we know $r+1$, and train the PC-LDA classifier on only the first $r+1$ principal components.  The approach described in Section \ref{subsec:CRC-L} in which all $n$ principal components are used would require a more detailed analysis.  Nonetheless, these observations are the basis for the statement in Section \ref{subsec:CRC-L} that when $\gamma$ is not sparse then CRC-L will capture the signal of interest and CRC-S may be redundant.  This also suggests that the case $p^+/p \to 0$ is the case of primary interest to us.

\section{Mean centering} \label{centering}

We do not include an intercept term in our model.  In practice, we may simply re-center the features to have mean 0.  More explicitly, suppose that we did include an intercept term $\mu$:
\[
	Z_{1 \times p} = \mu_{1 \times p} + T_{1 \times 1}\gamma_{1 \times p} + L_{1 \times r}\alpha_{r \times p} + \epsilon_{1 \times p}
\]
We could then estimate $\mu$ from the training data as $\hat{\mu} = n^{-1}\sum_{i=1}^n Z_i$ and let
\begin{align}
    \tilde{\bm{Z}} &= \bm{Z} - \bm{1}_n \hat{\mu} \label{center1} \\
    \tilde{Z} &= Z - \hat{\mu} \label{center2}
\end{align}
where $\bm{1}_n$ is a vector of 1's, and then use $\tilde{\bm{Z}}$ and $\tilde{Z}$ in place of $\bm{Z}$ and $Z$.

This introduces the complication that $\tilde{\bm{Z}}$ is not full rank and $\tilde{\bm{Z}}\tilde{\bm{Z}}'$ is therefore not invertible.  (Note that $\tilde{\bm{Z}} = (I - n^{-1}\bm{1}_n \bm{1}_n')\bm{Z}$ is the projection onto the orthogonal complement of $\bm{1}_n$.)  A simple fix is to replace the null eigenvalue of $\tilde{\bm{Z}}\tilde{\bm{Z}}'$ with some small value.  

In Appendix \ref{recovering_L} we note that, under the conditions given there, $\bm{Z}\bm{Z}'/p$ is approximately $\bm{L}\Lambda\bm{L}' + \sigma^2 I$.  In particular, the final $n - r$ eigenvalues of $\bm{Z}\bm{Z}'/p$ are all approximately $\sigma^2$.  If we were to replace $\hat{\mu}$ with $\mu$ in (\ref{center1}), the same argument would apply to $\bm{\tilde{Z}}\bm{\tilde{Z}}'/p$.  That is, if we were able to replace the estimate $\hat{\mu}$ with the true value $\mu$ in (\ref{center1}), then $\bm{\tilde{Z}}$ would be full rank, and the smallest $n-r$ eigenvalues of $\bm{\tilde{Z}}\bm{\tilde{Z}}'/p$ would all be approximately $\sigma^2$.  

This suggests that we replace the null eigenvalue of $\tilde{\bm{Z}}\tilde{\bm{Z}}'$ with a value roughly similar to the other smallest eigenvalues of $\tilde{\bm{Z}}\tilde{\bm{Z}}'$.  In practice, we use the median eigenvalue of $\tilde{\bm{Z}}\tilde{\bm{Z}}'$.

\section{Implementation of CRC-L} \label{crcl_implementation}

Let ``class 1'' denote the observations where $T_i = -1$ and let ``class 2'' denote the observations where $T_i = 1$.  Let $n_1$ and $n_2$ be the number of observations in class 1 and class 2, respectively.  Let $\bm{Y}_{n \times 2}$ be the class indicator matrix, i.e., $Y_{i1} = 1$ if $T_i$ = -1 and 0 otherwise, while $Y_{i2}$ = 1 if $T_i$ = 1 and 0 otherwise.

Let $\bm{Z} = \bm{UDV}'$ be the singular value decomposition of $\bm{Z}$, where $\bm{U}$ is $n \times n$, $\bm{D}$ is $n \times n$, and $\bm{V}$ is $n \times p$.  Then assuming $\bm{Z}$ has been mean centered (see Appendix \ref{centering}), $\bm{V}$ are the principal components of $\bm{Z}$.  Let $\bm{X} = \bm{Z}\bm{V}$ be the projection of $\bm{Z}$ onto $\bm{V}$ and similarly define $X = Z\bm{V}$ for the out-of-sample observation.  

Let $\hat{\mu}_1 = n_1^{-1}\sum_{i : T_i = -1} X_i$ and $\hat{\mu}_2 = n_2^{-1}\sum_{i : T_i = 1} X_i$ be the estimated class-conditional mean vectors.  Note that
\begin{equation*}
    \left( \begin{array}{c} \hat{\mu}_1 \\ \hat{\mu}_2 \end{array} \right) = (\bm{Y}'\bm{Y})^{-1}\bm{Y}'\bm{X}
\end{equation*}
and further define $d = (-1, 1)'$ so that $\hat{\mu}_2 - \hat{\mu}_1 = d'(\bm{Y}'\bm{Y})^{-1}\bm{Y}'\bm{X}$.

Let
\begin{align}
    \hat{\Sigma}_1 &= \frac{1}{n_1}\sum_{i:T_i=-1}(X_i - \hat{\mu}_1)'(X_i - \hat{\mu}_1) \label{sigma1}\\
    \hat{\Sigma}_2 &= \frac{1}{n_2}\sum_{i:T_i=1}(X_i - \hat{\mu}_2)'(X_i - \hat{\mu}_2) \label{sigma2}
\end{align}
and let 
\[ \hat{\Sigma} = (n_1/n) \hat{\Sigma}_1 + (n_2/n) \hat{\Sigma}_2. \]
% (Note the abuse of notation: $\hat{\Sigma}$ here is an $n \times n$ matrix and unrelated to the $p \times p$ matrix $\Sigma$ defined in Section 2.)
Alternatively $\hat{\Sigma}$ may be written as 
\begin{equation} \label{ldacovmat}
    \hat{\Sigma} = \frac{1}{n} \bm{X}'R_{\bm{Y}}\bm{X}
\end{equation}
where $R_{\bm{Y}} = I - \bm{Y}(\bm{Y}'\bm{Y})^{-1}\bm{Y}'$ is the residual operator of $\bm{Y}$.  Note that $\hat{\Sigma}$ is rank $n -2$, and therefore singular, due to the $R_{\bm{Y}}$ term in (\ref{ldacovmat}).  This issue is similar to the one discussed in Appendix \ref{centering}; in particular, the rank deficiency is due to the fact that the two class-conditional means $\hat{\mu}_1$ and $\hat{\mu}_2$ in (\ref{sigma1}) and (\ref{sigma2}) are estimated from $\bm{X}$.  Similarly to Appendix \ref{centering}, we therefore wish to modify $\hat{\Sigma}$ to make it invertible, and in particular replace the two null eigenvalues with some small value. 

The nullspace of $\hat{\Sigma}$ is spanned by $\bm{X}^{-1}\bm{Y}$; to see this, note $(\bm{X}'R_{\bm{Y}}\bm{X})(\bm{X}^{-1}\bm{Y}) = \bm{X}'R_{\bm{Y}}\bm{Y} = 0$.  We therefore form the projection matrix onto $\bm{X}^{-1}\bm{Y}$
\begin{align*}
    P &= (\bm{X}^{-1}\bm{Y}) \left[(\bm{X}^{-1}\bm{Y})'(\bm{X}^{-1}\bm{Y}) \right]^{-1} (\bm{X}^{-1}\bm{Y})' \\
    &= \bm{X}^{-1}\bm{Y} \left[\bm{Y}'(\bm{X}\bm{X}')^{-1}\bm{Y} \right]^{-1} \bm{Y}'\bm{X}^{-T}
\end{align*}
which has two eigenvalues that are 1, associated with two eigenvectors that span the nullspace of $\hat{\Sigma}$, and all remaining eigenvalues 0, and then define the augmented covariance matrix 
\begin{equation*} 
    \tilde{\Sigma} = \hat{\Sigma} + \lambda P
\end{equation*}
which has the same eigendecomposition as $\hat{\Sigma}$ except that the two null eigenvalues have been replaced by $\lambda$.   In practice, as in Appendix \ref{centering}, we set $\lambda$ equal to the median eigenvalue.

We may now construct an LDA classifier.  The score for an out-of-sample observation would be 
\begin{equation*} 
    s = X \tilde{\Sigma}^{-1} (\hat{\mu}_2 - \hat{\mu}_1)' - c 
\end{equation*}
where $c$ is a constant that we ignore (because we use the scores as an input to the ensemble classifier, $c$ is irrelevant).  We now note 
\begin{align*}
    X \tilde{\Sigma}^{-1} (\hat{\mu}_2 - \hat{\mu}_1) &= 
    X\left[\hat{\Sigma} + \lambda P \right]^{-1} \bm{X}'\bm{Y}(\bm{Y}'\bm{Y})^{-1}d \\
    &= Z\bm{V}\left[\hat{\Sigma} + \lambda P \right]^{-1} \bm{V}'\bm{Z}'\bm{Y}(\bm{Y}'\bm{Y})^{-1}d\\
    &= Z\bm{Z}'\bm{U}\bm{D}^{-1}\left[\hat{\Sigma} + \lambda P \right]^{-1} \bm{D} \bm{U}'\bm{Y}(\bm{Y}'\bm{Y})^{-1}d \\
    &= Z\bm{Z}'\left[\bm{U}\bm{D}^{-1}\hat{\Sigma}\bm{D}\bm{U}' + \lambda \bm{U}\bm{D}^{-1} P \bm{D}\bm{U}'\right]^{-1} \bm{Y}(\bm{Y}'\bm{Y})^{-1}d.    
\end{align*}
Moreover,
\begin{align*}
    \bm{U}\bm{D}^{-1}\hat{\Sigma}\bm{D}\bm{U}' &=
    \frac{1}{n}\bm{U}\bm{D}^{-1} \bm{X}'R_{\bm{Y}}\bm{X} \bm{D}\bm{U}' \\
    &= \frac{1}{n}\bm{U}\bm{D}^{-1} \bm{D}\bm{U}' R_{\bm{Y}} \bm{U}\bm{D} \bm{D}\bm{U}' \\
    &= \frac{1}{n} R_{\bm{Y}} \bm{U}\bm{D}^2\bm{U}'\\
    &= \frac{1}{n} R_{\bm{Y}} \bm{Z}\bm{Z}'
\end{align*}
and 
\begin{align*}
    \lambda \bm{U}\bm{D}^{-1}P\bm{D}\bm{U}' &=
    \lambda \bm{U}\bm{D}^{-1}   \bm{X}^{-1}\bm{Y} \left[\bm{Y}'(\bm{X}\bm{X}')^{-1}\bm{Y} \right]^{-1} \bm{Y}'\bm{X}^{-T}   \bm{D}\bm{U}' \\
    &= \lambda \bm{U}\bm{D}^{-1}   \bm{D}^{-1} \bm{U}' \bm{Y} \left[\bm{Y}'(\bm{Z}\bm{Z}')^{-1}\bm{Y} \right]^{-1} \bm{Y}'\bm{U} \bm{D}^{-1}   \bm{D}\bm{U}' \\
    &= \lambda (\bm{Z}\bm{Z}')^{-1} \bm{Y} \left[\bm{Y}'(\bm{Z}\bm{Z}')^{-1}\bm{Y} \right]^{-1} \bm{Y}' \\    
\end{align*}
and thus, putting this all together, the score (ignoring $c$) is
\begin{equation*}
    s = 
    Z\bm{Z}'\left\{\frac{1}{n} R_{\bm{Y}} \bm{Z}\bm{Z}' + \lambda  (\bm{Z}\bm{Z}')^{-1} \bm{Y} \left[\bm{Y}'(\bm{Z}\bm{Z}')^{-1}\bm{Y} \right]^{-1} \bm{Y}' \right\}^{-1} \bm{Y}(\bm{Y}'\bm{Y})^{-1}d.    
\end{equation*}
This is relatively straightforward to compute.  It involves computing the inverse of $n \times n$ matrices but does not require computing any $p \times p$ matrices.  Moreover, it is reasonably straightforward to obtain downdate formulas for use in leave-one-out calculations.  Note that when obtaining leave-one-out downdates, we use the same $\lambda$ throughout (i.e., we do not update $\lambda$ for each $i$), so that we only need to compute one eigendecomposition.

\section{Comment on $\hat{\gamma}$} \label{sec:gammahat}

The connection between $\bm{L}$ and $\bm{ZZ}'/p$ in Appendix \ref{recovering_L} lends insight into the estimate of $\gamma$ that we use in the residualization algorithm.  First, suppose momentarily that $\bm{L}$ were observed so that we could estimate $\gamma$ by ordinary least squares (OLS), regressing $\bm{Z}$ onto $\bm{T}$ and $\bm{L}$.  By the Frisch-Waugh-Lovell theorem, the OLS estimate of $\gamma$ obtained from this regression would be equal to 
\begin{equation}
    \hat{\gamma}^{OLS} = \left(\bm{T}' R_{\bm{L}} \bm{T}\right)^{-1}\bm{T}'R_{\bm{L}}\bm{Z} \label{gammaols}
\end{equation}
where $R_{\bm{L}} = I - \bm{L}\inv{(\bm{L}'\bm{L})}\bm{L}'$ denotes the orthogonal projection operator of $\bm{L}.$  

Now, observe that under the conditions specified in Appendix \ref{recovering_L}, we have
\begin{align*}
	\left(\frac{\bm{ZZ}'}{p \sigma^2}\right)^{-1} &{\overset{p}{\to}{}}
	\sigma^2 \left(\bm{L}\Lambda\bm{L}' + \sigma^2I\right)^{-1} \\
	&= I - \bm{L}(\bm{L}'\bm{L} + \sigma^2 \Lambda^{-1})^{-1}\bm{L}'
\end{align*}
as $p \rightarrow \infty.$ Define the matrix
\[
	\tilde{R}_{\bm{L}} = I - \bm{L}(\bm{L}'\bm{L} + \sigma^2 \Lambda^{-1})^{-1}\bm{L}'.
\]
Since $\bm{L}'\bm{L}$ scales with $n$ but $\sigma^2 \Lambda^{-1}$ does not, the $\sigma^2 \Lambda^{-1}$ term is negligible for large $n$ and  therefore $\tilde{R}_{\bm{L}} \approx R_{\bm{L}}$.

We can use $\tilde{R}_{\bm{L}}$ to rewrite $\hat{\gamma}$ from Section \ref{subsec:residualization} as
\begin{align*}
	\hat{\gamma}
	&= \left[\bm{T}' \left( \frac{\bm{ZZ}'}{p \sigma^2} \right)^{-1}\bm{T}\right]^{-1}\bm{T}'\left(\frac{\bm{ZZ}'}{p\sigma^2} \right)^{-1}\bm{Z} \\
	&= \left(\bm{T}' \tilde{R}_{\bm{L}}\bm{T}\right)^{-1}\bm{T}'\tilde{R}_{\bm{L}}\bm{Z} 
\end{align*}
and therefore $\hat{\gamma}$ can be viewed as approximating the OLS estimate (\ref{gammaols}).

\section{Feature selection}\label{feature_selection}

We tune $N,$ the number of `top' features to include in CRC-S, via a grid search that seeks to maximize the accuracy of the CRC. In our simulations and examples, we use the following grid: $(2^0, 2^{0.5}, 2^1, 2^{1.5}, \ldots, 2^{\lfloor \sqrt{p} \rfloor}).$

The grid search proceeds as follows. For each candidate value of $N,$ we construct $\hat{c}$ using only those features with the $N$ smallest $p$-values to construct CRC-S. We then estimate the error rate $e$ of $\hat{c}$ using the formula $$\hat{e} = 1-\Phi\left[\sqrt{(\hat{\mu}_2 - \hat{\mu}_1 )'\hat{\Sigma}^{-1} (\hat{\mu}_2 - \hat{\mu}_1)}\right]$$ where $\hat{\mu}_1$, $\hat{\mu}_2$, and $\hat{\Sigma}$ are the estimated class means and covariance matrix used in constructing $\hat{c}$. We then choose $N$ to be the candidate value which yields the minimum estimated error rate.
    
Although this procedure strongly resembles cross-validation, it is not a genuine cross-validation. This is because $(Z_i, T_i)$ was used in the cross-residualization of $\bm{Z}.$ As a result, $(Z_i, T_i)$ shows up implicitly in $\drop{\bm{S}}.$ Since we use $\drop{\bm{S}}$ to predict $T_i$, this may lead to overfitting in the DLDA classifier, particularly for large candidate values of $N$. To alleviate the problem, we project out the $i$th row of $(\bm{ZZ}')^{-1}\bm{Z}$ from the class means computed from $\drop{\hat{\bm{S}}}$, which has the effect of removing the contribution of $Z_i$ to $\drop{\hat{\bm{S}}}$.

\end{appendix}

\FloatBarrier

\section*{Acknowledgements}
NYP was supported by a NSF Graduate Research Fellowship (DGE 1256260). This research was supported in part through computational resources and services provided by Advanced Research Computing at the University of Michigan, Ann Arbor.
The authors would like to thank Zhihao Guo, Greg Hunt, Kristen Hunter, Dan Kessler, Zoe Rehnberg, Kerby Shedden, Terry Speed, and Jonathan Terhorst for helpful comments.

\FloatBarrier

\section*{Supplement}

The supplement contains three sections. Section S1 contains additional simulation results. Section S2 gives additional details about the genomic datasets used in the paper, including preprocessing steps performed. Section S3 contains additional analysis results for the genomic datasets.

\setcounter{subsection}{0}
\renewcommand{\thesubsection}{S\arabic{subsection}}
\setcounter{table}{0}
\renewcommand{\thetable}{S\arabic{table}}
\setcounter{figure}{0}
\renewcommand{\thefigure}{S\arabic{figure}}
\setcounter{figure}{0}

\subsection{Additional simulation results} \label{sim_supplement}

Due to computational constraints, we vary the number of replications with $n:$ 100 replications are performed for $n=50$, $100$, and $200;$ 30 replications for $n=500,$ and 10 replications for $n=1000.$

In the main text, we present simulations for $p=\bignumber{100000}.$ Here we give additional results for $p=\bignumber{20000}$ and $p=\bignumber{500000},$ so that we may examine the effect of varying $p$ in addition to the effect of varying $n.$ 

CRC-S, CRC-L, and CRC exhibit largely the same performance for the different values of $p$ we examined. At small sample sizes, accuracy is degraded as $p$ increases, but we see this with the other classifiers as well. Comparing CRC to other classifiers, we see that as $p$ increases, the relative gap in accuracy between CRC and glmnet increases slightly (particularly for large $n$).

\begin{figure}[h]
    \centering
    \includegraphics[width=.8\linewidth]{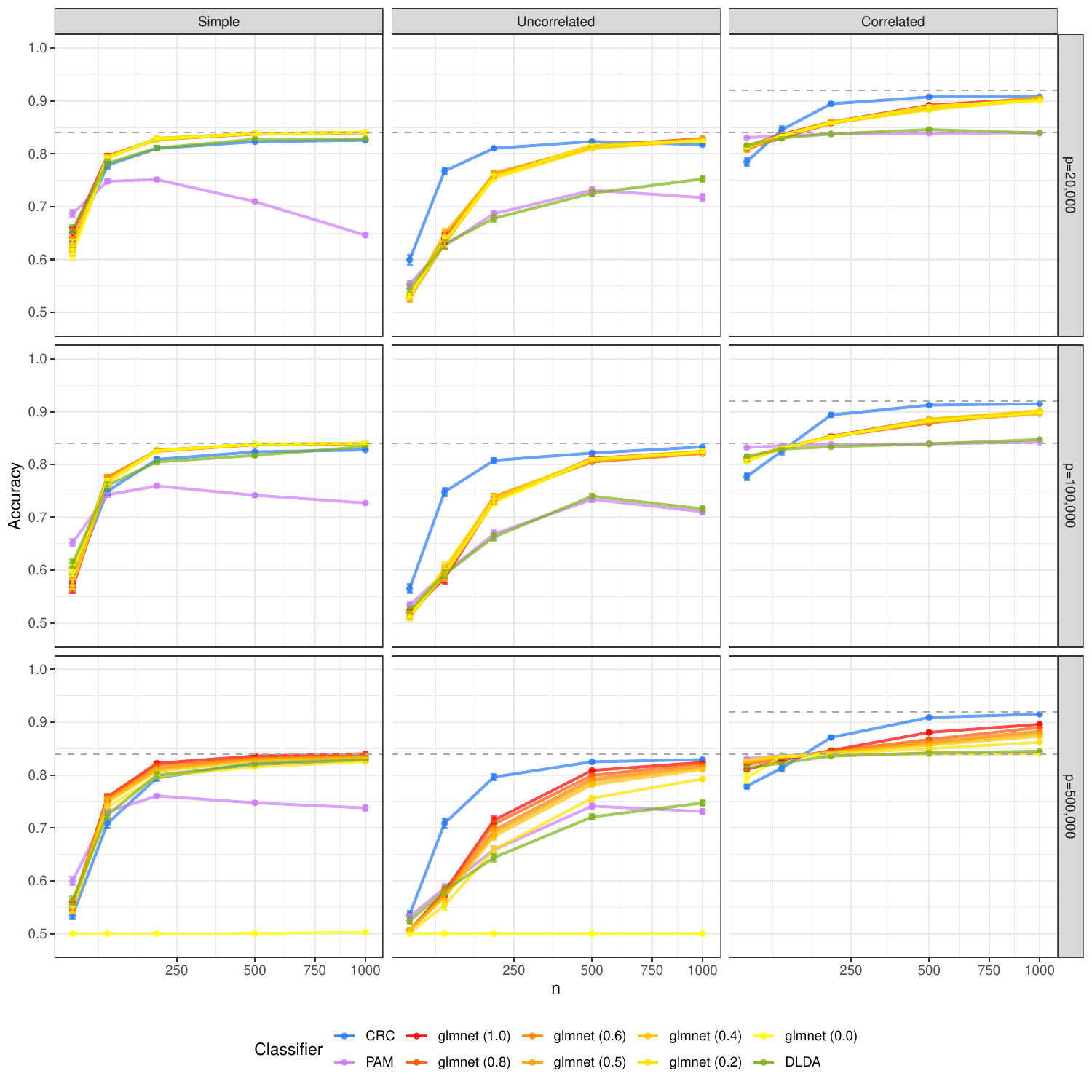}
    \label{fig:all_clfs_grid}
    \caption{Mean accuracies of glmnet, PAM, DLDA, and CRC for $p=\bignumber{20000}$, $p=\bignumber{100000}$, and $p=\bignumber{500000}.$ Sample size $n$ is depicted on a square-root scale. The dashed horizontal lines are at $\Phi(1)$ and $\Phi(\sqrt{2})$ and indicate Bayes optimal accuracy rates (as detailed in the main text). For glmnet, the parameter $\alpha$ is noted in parentheses.}
\end{figure}

For smaller $p,$ the $\alpha$ regularization parameter in glmnet does not seem to affect its accuracy very much (see the overlapping curves for glmnet in Figure  \ref{fig:all_clfs_grid}). However, for large $p,$ this parameter in glmnet becomes important, particularly in the uncorrelated and correlated cases where the difference between glmnet for $\alpha=1$ and smaller values of $\alpha$ is quite large.

\begin{figure}[h]
    \centering
    \includegraphics[width=.8\linewidth]{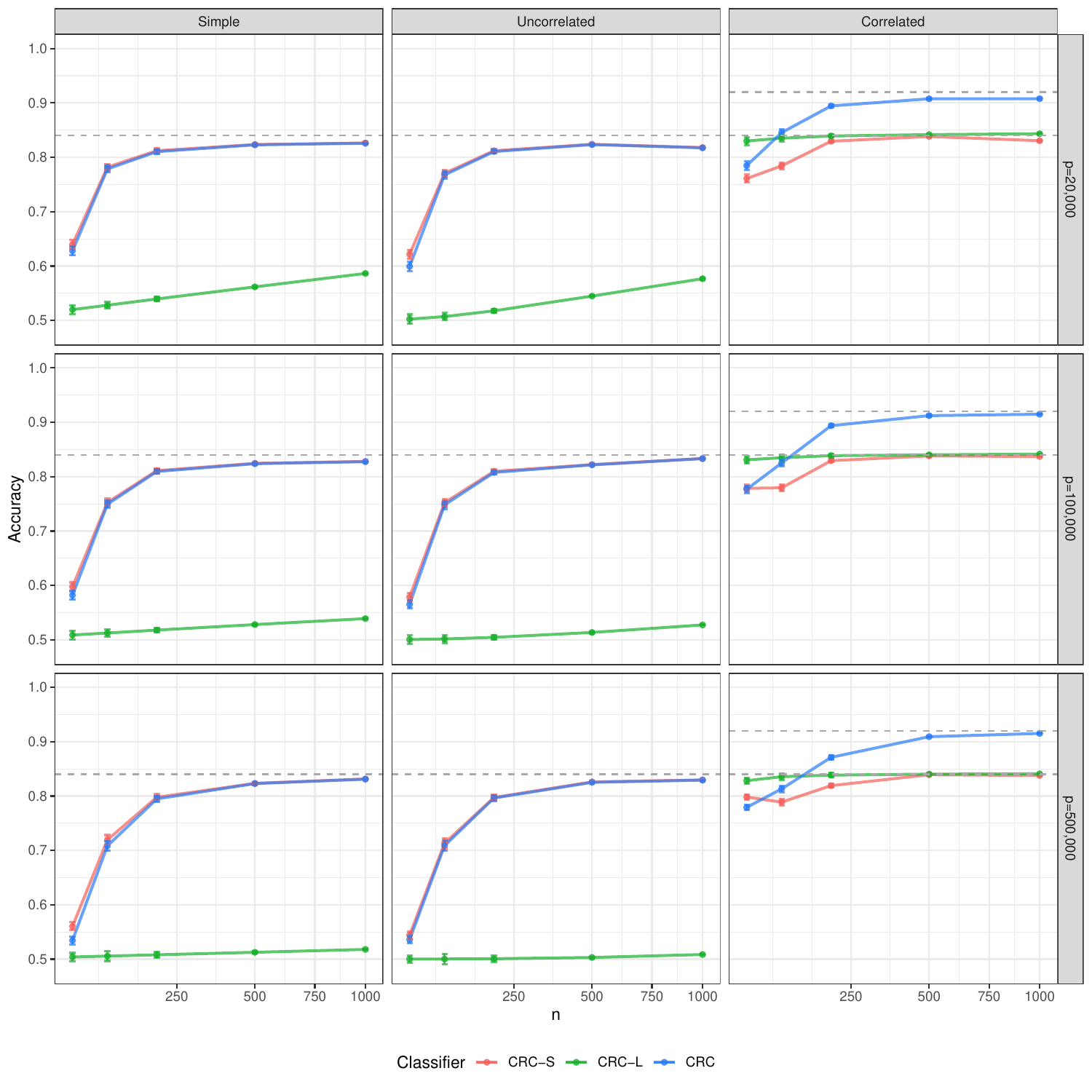}
    \caption{Mean accuracies of CRC, CRC-S, and CRC-L  for $p=\bignumber{20000}$, $p=\bignumber{100000}$, and $p=\bignumber{500000}$. Sample size $n$ is depicted on a square-root scale. The dashed horizontal lines are at $\Phi(1)$ and $\Phi(\sqrt{2})$ and indicate Bayes optimal accuracy rates (as detailed in the main text).}
\end{figure}

\FloatBarrier

%%%%%%%%%%%%%%%%%%%%%%%%%%%%%%%%%%%%%%%%%%%%%%%%%%%%%%%%%%%%%%%
\FloatBarrier
\subsection{Datasets}\label{supp:datasets}

All data were downloaded from GEO, with the exception of the colorectal cancer dataset, which was downloaded from Array Express. Accession numbers and the associated studies are cited in Table \ref{tab:accession_studies}.
The datasets were pre-processed as follows.

\emph{Microarray data:}
Microarray data were background corrected using robust multi-array averaging as implemented in the R package \texttt{oligo} \citep{irizarry2003exploration,oligo}.

\emph{RNA-seq data:}
For RNA-seq data, we applied a $\log_2(x+1)$ transform to the count matrix, after identifying and removing duplicated features. We accounted for library size by centering the rows of the count matrix (i.e., observations). Note that there is a fourth class, ``Cancer,'' in the full Sepsis dataset. We omit the cancer samples because the number of samples for this patient group is small (14 samples across 5 patients).

\emph{Methylation data:}
For the Crohn's disease methylation data, we utilize the  beta-values deposited on GEO, which have already been background-corrected. For all other methylation datasets, we download the raw files and  perform background correction using the R package \texttt{minfi} \citep{minfi}. Several features in the Crohn's disease methylation dataset have missing values. We remove these features prior to analysis. Additionally, this dataset contains a mixture of baseline and follow-up observations. We use only the baseline observations in our analysis.

Two datasets in our analysis (Sepsis and Alzheimer's) contain multiple samples per individual. When creating train-test splits, we make sure that samples belonging to a given individual are not split across the training and test sets. More specifically, we use the same splitting procedure we use for the other datasets, except now we partition individuals rather than samples. If an individual is selected for inclusion in the training set, we include all samples associated with that individual in the training set (similarly if selected for inclusion in the test set).

\begin{table}[h]
\caption{Accession numbers and associated studies. Datasets with an accession number starting with `GSE' are from GEO, datasets with an accession number starting with `E-MTAB' are from ArrayExpress.}
\label{tab:accession_studies}
\centering
\vspace{0.2cm}
\begin{tabular}{lll}
\toprule
Dataset           & Accession Number & Study                                    \\
\midrule
Alzheimer's       & GSE 66351        & \cite{gasparoni2018dna} \\
Asthma            & GSE 85566        & \cite{nicodemus2016dna} \\
Crohn's (Expr.)   & GSE 101794       & \cite{haberman2019age}                 \\
Crohn's (Methyl.) & GSE 112611       & \cite{somineni2019blood}                 \\
Colorectal Cancer & E-MTAB 1532      & \cite{xu2013identification}                 \\
FASD              & GSE 112987       & \cite{cobben2018dna}                 \\
Sepsis            & GSE 133822       & \cite{washburn2019t}                \\
\bottomrule
\end{tabular}
\end{table}

%%%%%%%%%%%%%%%%%%%%%%%%%%%%%%%%%%%%%%%%%%%%%%%%%%%%%%%%%%%%%%%
\FloatBarrier
\subsection{Additional Analysis Results}\label{supp:additional_analysis}

\begin{figure}[h]
\centering
\includegraphics[width=.8\linewidth]{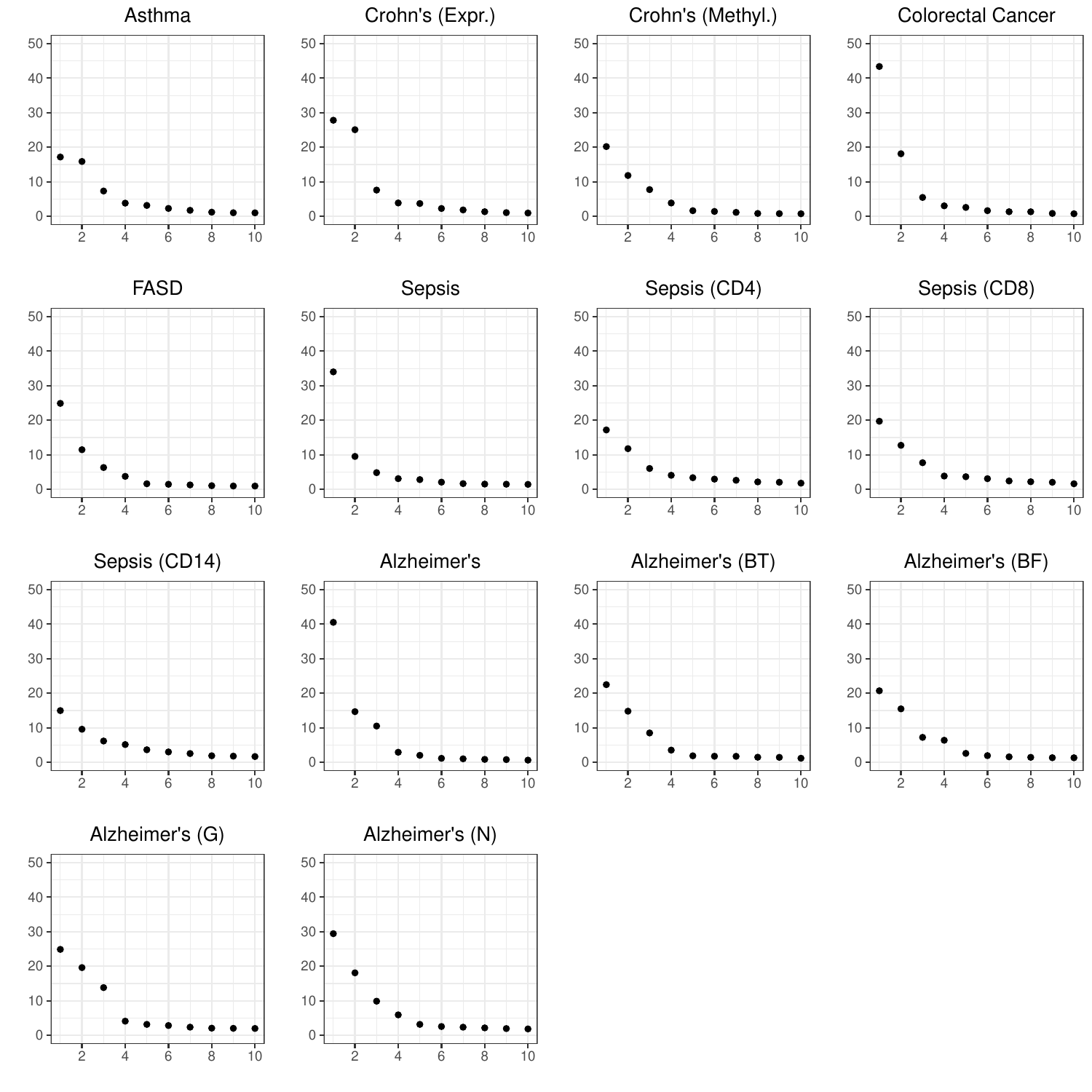}
\caption{Scree plots of the datasets, depicting percentage of variance explained by the first ten principal components.}
\label{scree_plots}
\end{figure}

Scree plots for the datasets in Table \ref{table:datasets} are given in Figure \ref{scree_plots}. Since the Alzheimer's and Sepsis datasets are comprised of various cell and tissue types, we also show scree plots for each subgroup. Note that in all datasets a large portion of the variance in the data is explained by the first ten principal components.

In the CRC, we do not explicitly estimate $r,$ the rank of $\bm{L},$ choosing instead to leverage the properties of ridgeless regression. If instead we were to estimate $r,$ one way would be to use the scree plots to locate the point at which the variance explained drops sharply (the `elbow'). This point is more easily located for some datasets than others, motivating our choice to avoid explicit estimation of $r.$

\FloatBarrier
The Sepsis and Alzheimer's datasets contain samples of various types, which define several subgroups of interest. Here we report within-subgroup accuracies for the analysis performed in Section \ref{sec:real-data}. In particular, the accuracies reported below are not from new analyses in which we have subsetted our data to the subgroup of interest, but rather the classifiers are trained on observations of all sample types as in Section \ref{sec:real-data}. The CRC generally performs as well or better than the other classifiers within the various sample types. The improvement within subgroups can be sizeable, as we see in the Sepsis data (Critically Ill vs. Sepsis, Critically Ill vs. Healthy).

\begin{table}[h]
\centering
\caption{Accuracies within sample type for the Alzheimer's data. Header abbreviations: BF: bulk samples from frontal cortex, BT: bulk samples from temporal cortex, G: purified glia samples, N: purified neuron samples.}
\label{tab:alz_cell_type}
\vspace{0.2cm}
\scalebox{0.9}{
\begin{tabular}{rrrrrrrrrrrrrrrr}
\toprule
\multicolumn{4}{c}{glmnet ($\alpha = 1$)} & \multicolumn{4}{c}{PAM} & \multicolumn{4}{c}{DLDA} & \multicolumn{4}{c}{CRC} \\
\cmidrule(l{3pt}r{3pt}){1-4} \cmidrule(l{3pt}r{3pt}){5-8} \cmidrule(l{3pt}r{3pt}){9-12} \cmidrule(l{3pt}r{3pt}){13-16}
BF & BT & G & N & BF & BT & G & N & BF & BT & G & N & BF & BT & G & N\\
\midrule
0.70 & 0.74 & 0.8 & 0.76 & 0.71 & 0.71 & 0.80 & 0.66 & 0.71 & 0.7 & 0.77 & 0.61 & 0.76 & 0.78 & 0.81 & 0.80\\
\bottomrule
\end{tabular}
}
\end{table}

\begin{table}[h]
\centering
\caption{Accuracies within sample type for the Sepsis data. Abbreviations: H: Healthy, CI: Critically Ill, S: Sepsis.}
\label{tab:sepsis_cell_type}
\vspace{0.2cm}
\scalebox{0.9}{
\begin{tabular}{lrrrrrrrrrrrr}
\toprule
\multicolumn{1}{c}{ } & \multicolumn{3}{c}{glmnet ($\alpha = 1$)} & \multicolumn{3}{c}{PAM} & \multicolumn{3}{c}{DLDA} & \multicolumn{3}{c}{CRC} \\
\cmidrule(l{3pt}r{3pt}){2-4} \cmidrule(l{3pt}r{3pt}){5-7} \cmidrule(l{3pt}r{3pt}){8-10} \cmidrule(l{3pt}r{3pt}){11-13}
  & CD14 & CD4 & CD8 & CD14 & CD4 & CD8 & CD14 & CD4 & CD8 & CD14 & CD4 & CD8\\
\midrule
H vs. S & 0.97 & 0.94 & 0.93 & 0.95 & 0.92 & 0.83 & 0.95 & 0.93 & 0.84 & 0.96 & 0.95 & 0.94\\
CI vs. S & 0.72 & 0.71 & 0.67 & 0.67 & 0.74 & 0.69 & 0.68 & 0.70 & 0.69 & 0.80 & 0.78 & 0.74\\
CI vs. H & 0.89 & 0.84 & 0.82 & 0.86 & 0.81 & 0.74 & 0.86 & 0.86 & 0.69 & 0.91 & 0.90 & 0.86\\
\bottomrule
\end{tabular}
}
\end{table}

\FloatBarrier
Tables \ref{tab:results_SE}, \ref{tab:alz_SE}, and \ref{tab:sepsis_SE} contain standard error estimates for the mean test accuracies reported in Tables \ref{tab:results}, \ref{tab:alz_cell_type}, and \ref{tab:sepsis_cell_type} respectively. These standard errors are calculated by taking the standard deviation of the mean test accuracies (resulting from the 200 train-test splits) and dividing by $\sqrt{200}.$

\begin{table}[h]
\caption{Standard errors for accuracies in Table \ref{tab:results}.}
\label{tab:results_SE}
\vspace{0.2cm}
\centering
\begin{tabular}{lrrrrrr}
\toprule
  & glmnet & PAM & DLDA & CRC & CRC-S & CRC-L\\
\midrule
Alzheimer's & 0.006 & 0.007 & 0.007 & 0.006 & 0.006 & 0.006\\
Asthma & 0.008 & 0.006 & 0.007 & 0.008 & 0.008 & 0.007\\
Colorectal Cancer & 0.003 & 0.006 & 0.005 & 0.003 & 0.003 & 0.004\\
Crohn's (Expression) & 0.004 & 0.005 & 0.005 & 0.004 & 0.006 & 0.004\\
Crohn's (Methylation) & 0.004 & 0.004 & 0.004 & 0.004 & 0.004 & 0.005\\
FASD & 0.007 & 0.007 & 0.007 & 0.008 & 0.008 & 0.008\\
Sepsis (Healthy vs. Sepsis) & 0.004 & 0.005 & 0.005 & 0.004 & 0.004 & 0.005\\
Sepsis (Crit. Ill. vs. Sepsis) & 0.008 & 0.007 & 0.008 & 0.008 & 0.008 & 0.007\\
Sepsis (Crit. Ill. vs. Healthy) & 0.006 & 0.006 & 0.006 & 0.006 & 0.006 & 0.007\\
\bottomrule
\end{tabular}
\end{table}

\begin{table}[h]
\centering
\caption{Standard errors for accuracy rates in Table \ref{tab:alz_cell_type}.}
\label{tab:alz_SE}
\vspace{0.2cm}
\begin{tabular}{rrrrrrrrrrrrrrrr}
\toprule
\multicolumn{4}{c}{glmnet ($\alpha = 1$)} & \multicolumn{4}{c}{PAM}\\
\cmidrule(l{3pt}r{3pt}){1-4} \cmidrule(l{3pt}r{3pt}){5-8}
BF & BT & G & N & BF & BT & G & N \\
\midrule
0.009 & 0.009 & 0.013 & 0.013 & 0.01 & 0.009 & 0.012 & 0.015 \\
\bottomrule
\end{tabular}

\begin{tabular}{rrrrrrrrrrrrrrrr}
\toprule
\multicolumn{4}{c}{DLDA} & \multicolumn{4}{c}{CRC} \\
\cmidrule(l{3pt}r{3pt}){1-4} \cmidrule(l{3pt}r{3pt}){5-8} 
BF & BT & G & N & BF & BT & G & N \\
\midrule
0.009 & 0.010 & 0.013 & 0.014 & 0.008 & 0.008 & 0.013 & 0.013\\
\bottomrule
\end{tabular}
\end{table}

\begin{table}[h]
\centering
\caption{Standard errors for accuracy rates in Table \ref{tab:sepsis_cell_type}.}
\label{tab:sepsis_SE}
\vspace{0.2cm}
\begin{tabular}{lrrrrrr}
\toprule
\multicolumn{1}{c}{ } & \multicolumn{3}{c}{glmnet ($\alpha = 1$)} & \multicolumn{3}{c}{PAM} \\
\cmidrule(l{3pt}r{3pt}){2-4} \cmidrule(l{3pt}r{3pt}){5-7}
  & CD14 & CD4 & CD8 & CD14 & CD4 & CD8\\
\midrule
H vs. S & 0.004 & 0.005 & 0.006 & 0.005 & 0.006 & 0.009\\
CI vs. S & 0.010 & 0.010 & 0.010 & 0.009 & 0.010 & 0.009\\
CI vs. H & 0.007 & 0.009 & 0.009 & 0.008 & 0.009 & 0.008\\
\bottomrule
\end{tabular}

\begin{tabular}{lrrrrrr}
\toprule
\multicolumn{1}{c}{ } & \multicolumn{3}{c}{DLDA} & \multicolumn{3}{c}{CRC} \\
\cmidrule(l{3pt}r{3pt}){2-4} \cmidrule(l{3pt}r{3pt}){5-7}
  & CD14 & CD4 & CD8 & CD14 & CD4 & CD8\\
\midrule
H vs. S & 0.005 & 0.007 & 0.009 & 0.004 & 0.005 & 0.005\\
CI vs. S & 0.009 & 0.012 & 0.011 & 0.009 & 0.010 & 0.011\\
CI vs. H & 0.009 & 0.010 & 0.009 & 0.007 & 0.007 & 0.008\\
\bottomrule
\end{tabular}
\end{table}

\FloatBarrier

\bibliographystyle{abbrvnat}
\bibliography{refs}

\begin{thebibliography}{56}
\providecommand{\natexlab}[1]{#1}
\providecommand{\url}[1]{\texttt{#1}}
\expandafter\ifx\csname urlstyle\endcsname\relax
  \providecommand{\doi}[1]{doi: #1}\else
  \providecommand{\doi}{doi: \begingroup \urlstyle{rm}\Url}\fi

\bibitem[Aryee et~al.(2014)Aryee, Jaffe, Corrada-Bravo, Ladd-Acosta, Feinberg,
  Hansen, and Irizarry]{minfi}
M.~J. Aryee, A.~E. Jaffe, H.~Corrada-Bravo, C.~Ladd-Acosta, A.~P. Feinberg,
  K.~D. Hansen, and R.~A. Irizarry.
\newblock {Minfi: A flexible and comprehensive Bioconductor package for the
  analysis of Infinium DNA Methylation microarrays}.
\newblock \emph{Bioinformatics}, 30\penalty0 (10):\penalty0 1363--1369, 2014.
\newblock \doi{10.1093/bioinformatics/btu049}.

\bibitem[Bickel and Levina(2004)]{bickel2004some}
P.~J. Bickel and E.~Levina.
\newblock Some theory for {F}isher's linear discriminant function, `naive
  {B}ayes', and some alternatives when there are many more variables than
  observations.
\newblock \emph{Bernoulli}, 10\penalty0 (6):\penalty0 989--1010, 2004.

\bibitem[Bind et~al.(2014)Bind, Lepeule, Zanobetti, Gasparrini, Baccarelli,
  Coull, Tarantini, Vokonas, Koutrakis, and Schwartz]{bind2014air}
M.-A. Bind, J.~Lepeule, A.~Zanobetti, A.~Gasparrini, A.~A. Baccarelli, B.~A.
  Coull, L.~Tarantini, P.~S. Vokonas, P.~Koutrakis, and J.~Schwartz.
\newblock Air pollution and gene-specific methylation in the {N}ormative
  {A}ging {S}tudy: Association, effect modification, and mediation analysis.
\newblock \emph{Epigenetics}, 9\penalty0 (3):\penalty0 448--458, 2014.

\bibitem[Bollepalli et~al.(2019)Bollepalli, Korhonen, Kaprio, Anders, and
  Ollikainen]{bollepalli2019epismoker}
S.~Bollepalli, T.~Korhonen, J.~Kaprio, S.~Anders, and M.~Ollikainen.
\newblock Epismoker: A robust classifier to determine smoking status from {DNA}
  methylation data.
\newblock \emph{Epigenomics}, 11\penalty0 (13):\penalty0 1469--1486, 2019.

\bibitem[Boyle et~al.(2017)Boyle, Li, and Pritchard]{boyle2017expanded}
E.~A. Boyle, Y.~I. Li, and J.~K. Pritchard.
\newblock An expanded view of complex traits: From polygenic to omnigenic.
\newblock \emph{Cell}, 169\penalty0 (7):\penalty0 1177--1186, 2017.

\bibitem[Carvalho and Irizarry(2010)]{oligo}
B.~S. Carvalho and R.~A. Irizarry.
\newblock A framework for oligonucleotide microarray preprocessing.
\newblock \emph{Bioinformatics}, 26\penalty0 (19):\penalty0 2363--7, 2010.
\newblock ISSN 1367-4803.
\newblock \doi{10.1093/bioinformatics/btq431}.

\bibitem[Choi et~al.(2017)Choi, Taylor, and Tibshirani]{choi2017selecting}
Y.~Choi, J.~Taylor, and R.~Tibshirani.
\newblock Selecting the number of principal components: Estimation of the true
  rank of a noisy matrix.
\newblock \emph{Annals of Statistics}, 45\penalty0 (6):\penalty0 2590--2617,
  2017.

\bibitem[Cobben et~al.(2018)Cobben, Krzyzewska, Venema, Mul, Polstra, Postma,
  Smigiel, Pesz, Niklinski, Chomczyk, Henneman, and Mannens]{cobben2018dna}
J.~M. Cobben, I.~M. Krzyzewska, A.~Venema, A.~N. Mul, A.~Polstra, A.~V. Postma,
  R.~Smigiel, K.~Pesz, J.~Niklinski, M.~A. Chomczyk, P.~Henneman, and M.~M.
  A.~M. Mannens.
\newblock {DNA} methylation abundantly associates with fetal alcohol spectrum
  disorder and its subphenotypes.
\newblock \emph{Epigenomics}, 11\penalty0 (7):\penalty0 767--785, 2018.

\bibitem[Cook et~al.(2012)Cook, Forzani, and Rothman]{cook2012estimating}
R.~D. Cook, L.~Forzani, and A.~J. Rothman.
\newblock Estimating sufficient reductions of the predictors in abundant
  high-dimensional regressions.
\newblock \emph{Annals of Statistics}, 40\penalty0 (1):\penalty0 353--384,
  2012.

\bibitem[Dicker(2012)]{dicker2012optimal}
L.~Dicker.
\newblock Optimal estimation and prediction for dense signals in
  high-dimensional linear models.
\newblock \emph{arXiv:1203.4572}, 2012.

\bibitem[Dobriban and Wager(2018)]{dobriban2018high}
E.~Dobriban and S.~Wager.
\newblock High-dimensional asymptotics of prediction: Ridge regression and
  classification.
\newblock \emph{Annals of Statistics}, 46\penalty0 (1):\penalty0 247--279,
  2018.

\bibitem[Dudoit and Fridlyand(2003)]{dudoit2003classification}
S.~Dudoit and J.~Fridlyand.
\newblock Classification in microarray experiments.
\newblock In T.~Speed, editor, \emph{Statistical Analysis of Gene Expression
  Microarray Data}, pages 93--158. Chapman \& Hall/CRC, 2003.

\bibitem[Elliott et~al.(2014)Elliott, Tillin, McArdle, Ho, Duggirala, Frayling,
  Smith, Hughes, Chaturvedi, and Relton]{elliott2014differences}
H.~R. Elliott, T.~Tillin, W.~L. McArdle, K.~Ho, A.~Duggirala, T.~M. Frayling,
  G.~D. Smith, A.~D. Hughes, N.~Chaturvedi, and C.~L. Relton.
\newblock Differences in smoking associated {DNA} methylation patterns in
  {S}outh {A}sians and {E}uropeans.
\newblock \emph{Clinical Epigenetics}, 6\penalty0 (1):\penalty0 4, 2014.

\bibitem[Fan and Fan(2008)]{fan2008high}
J.~Fan and Y.~Fan.
\newblock High dimensional classification using features annealed independence
  rules.
\newblock \emph{Annals of Statistics}, 36\penalty0 (6):\penalty0 2605, 2008.

\bibitem[Fan et~al.(2012)Fan, Feng, and Tong]{fan2012road}
J.~Fan, Y.~Feng, and X.~Tong.
\newblock A road to classification in high dimensional space: The regularized
  optimal affine discriminant.
\newblock \emph{Journal of the Royal Statistical Society: Series B},
  74\penalty0 (4):\penalty0 745--771, 2012.

\bibitem[Fan et~al.(2013)Fan, Liao, and Mincheva]{fan2013large}
J.~Fan, Y.~Liao, and M.~Mincheva.
\newblock Large covariance estimation by thresholding principal orthogonal
  complements.
\newblock \emph{Journal of the Royal Statistical Society: Series B},
  75\penalty0 (4):\penalty0 603--680, 2013.

\bibitem[Friedman et~al.(2001)Friedman, Hastie, and
  Tibshirani]{friedman2001elements}
J.~Friedman, T.~Hastie, and R.~Tibshirani.
\newblock \emph{The Elements of Statistical Learning}, volume~1.
\newblock Springer Series in Statistics, 2001.

\bibitem[Gagnon-Bartsch and Speed(2012)]{gagnonbartsch2012using}
J.~Gagnon-Bartsch and T.~Speed.
\newblock {Using control genes to correct for unwanted variation in microarray
  data}.
\newblock \emph{Biostatistics}, 13\penalty0 (3):\penalty0 539--552, 2012.

\bibitem[Gagnon-Bartsch et~al.(2013)Gagnon-Bartsch, Jacob, and
  Speed]{gagnon2013removing}
J.~A. Gagnon-Bartsch, L.~Jacob, and T.~P. Speed.
\newblock Removing unwanted variation from high dimensional data with negative
  controls.
\newblock Technical report, UC Berkeley Department of Statistics, 2013.

\bibitem[Gasparoni et~al.(2018)Gasparoni, Bultmann, Lutsik, Kraus, Sordon,
  Vlcek, Dietinger, Steinmaurer, Haider, Mulholland, Arzberger, Roeber,
  Riemenschneider, Kretzschmar, Giese, Leonhardt, and Walter]{gasparoni2018dna}
G.~Gasparoni, S.~Bultmann, P.~Lutsik, T.~F. Kraus, S.~Sordon, J.~Vlcek,
  V.~Dietinger, M.~Steinmaurer, M.~Haider, C.~B. Mulholland, T.~Arzberger,
  S.~Roeber, M.~Riemenschneider, H.~A. Kretzschmar, A.~Giese, H.~Leonhardt, and
  J.~Walter.
\newblock {DNA} methylation analysis on purified neurons and glia dissects age
  and {A}lzheimer's disease-specific changes in the human cortex.
\newblock \emph{Epigenetics \& Chromatin}, 11\penalty0 (1):\penalty0 41, 2018.

\bibitem[Gerard and Stephens(2021)]{gerard2021unifying}
D.~Gerard and M.~Stephens.
\newblock Unifying and generalizing methods for removing unwanted variation
  based on negative controls.
\newblock \emph{Statistica Sinica}, 31\penalty0 (3), 2021.
\newblock \doi{doi:10.5705/ss.202018.0345}.

\bibitem[Guyon and Elisseeff(2003)]{guyon2003introduction}
I.~Guyon and A.~Elisseeff.
\newblock An introduction to variable and feature selection.
\newblock \emph{Journal of Machine Learning Research}, 3:\penalty0 1157--1182,
  2003.

\bibitem[Haberman et~al.(2019)Haberman, Schirmer, Dexheimer, Karns, Braun, Kim,
  Walters, Baldassano, Noe, Rosh, Markowitz, Crandall, Mack, Griffiths, Heyman,
  Baker, Kellermayer, Moulton, Patel, Gulati, Steiner, LeLeiko, Otley,
  Oliva-Hemker, Ziring, Kirschner, Keljo, Guthery, Cohen, Snapper, Evans,
  Dubinsky, Aronow, Hyams, Kugathasan, Huttenhower, Xavier, and
  Denson]{haberman2019age}
Y.~Haberman, M.~Schirmer, P.~J. Dexheimer, R.~Karns, T.~Braun, M.-O. Kim, T.~D.
  Walters, R.~N. Baldassano, J.~D. Noe, J.~Rosh, J.~Markowitz, W.~Crandall,
  D.~Mack, A.~Griffiths, M.~Heyman, S.~Baker, R.~Kellermayer, D.~Moulton,
  A.~Patel, A.~Gulati, S.~Steiner, N.~LeLeiko, A.~Otley, M.~Oliva-Hemker,
  D.~Ziring, B.~Kirschner, D.~Keljo, S.~Guthery, S.~Cohen, S.~Snapper,
  J.~Evans, M.~Dubinsky, B.~Aronow, J.~Hyams, S.~Kugathasan, C.~Huttenhower,
  R.~Xavier, and L.~Denson.
\newblock Age-of-diagnosis dependent ileal immune intensification and reduced
  alpha-defensin in older versus younger pediatric {C}rohn {D}isease patients
  despite already established dysbiosis.
\newblock \emph{Mucosal Immunology}, 12\penalty0 (2):\penalty0 491--502, 2019.

\bibitem[Hall et~al.(2014)Hall, Jin, and Miller]{hall2014feature}
P.~Hall, J.~Jin, and H.~Miller.
\newblock Feature selection when there are many influential features.
\newblock \emph{Bernoulli}, 20\penalty0 (3):\penalty0 1647--1671, 2014.

\bibitem[Hastie et~al.(2019)Hastie, Montanari, Rosset, and
  Tibshirani]{hastie2019surprises}
T.~Hastie, A.~Montanari, S.~Rosset, and R.~J. Tibshirani.
\newblock Surprises in high-dimensional ridgeless least squares interpolation.
\newblock \emph{arXiv preprint arXiv:1903.08560}, 2019.

\bibitem[Irizarry et~al.(2003)Irizarry, Hobbs, Collin, Beazer-Barclay,
  Antonellis, Scherf, and Speed]{irizarry2003exploration}
R.~A. Irizarry, B.~Hobbs, F.~Collin, Y.~D. Beazer-Barclay, K.~J. Antonellis,
  U.~Scherf, and T.~P. Speed.
\newblock Exploration, normalization, and summaries of high density
  oligonucleotide array probe level data.
\newblock \emph{Biostatistics}, 4\penalty0 (2):\penalty0 249--264, 2003.

\bibitem[Jolliffe(1986)]{jolliffe1986principal}
I.~T. Jolliffe.
\newblock Principal components in regression analysis.
\newblock In \emph{Principal Component Analysis}, pages 129--155. Springer,
  1986.

\bibitem[Kneip and Sarda(2011)]{kneip2011factor}
A.~Kneip and P.~Sarda.
\newblock Factor models and variable selection in high-dimensional regression
  analysis.
\newblock \emph{Annals of Statistics}, 39\penalty0 (5):\penalty0 2410--2447,
  2011.

\bibitem[Lee and Pausova(2013)]{lee2013cigarette}
K.~W. Lee and Z.~Pausova.
\newblock Cigarette smoking and {DNA} methylation.
\newblock \emph{Frontiers in Genetics}, 4:\penalty0 132, 2013.

\bibitem[Leek and Storey(2008)]{leek2008general}
J.~Leek and J.~Storey.
\newblock {A general framework for multiple testing dependence}.
\newblock \emph{Proceedings of the National Academy of Sciences}, 105\penalty0
  (48):\penalty0 18718--18723, 2008.
\newblock ISSN 0027-8424.

\bibitem[Leek and Storey(2007)]{leek2007capturing}
J.~T. Leek and J.~D. Storey.
\newblock Capturing heterogeneity in gene expression studies by surrogate
  variable analysis.
\newblock \emph{PLoS Genetics}, 3\penalty0 (9):\penalty0 e161, 2007.

\bibitem[Listgarten et~al.(2010)Listgarten, Kadie, Schadt, and
  Heckerman]{listgarten2010correction}
J.~Listgarten, C.~Kadie, E.~Schadt, and D.~Heckerman.
\newblock {Correction for hidden confounders in the genetic analysis of gene
  expression}.
\newblock \emph{Proceedings of the National Academy of Sciences}, 107\penalty0
  (38):\penalty0 16465, 2010.
\newblock ISSN 0027-8424.

\bibitem[Nguyen and Rocke(2002)]{nguyen2002tumor}
D.~V. Nguyen and D.~M. Rocke.
\newblock Tumor classification by partial least squares using microarray gene
  expression data.
\newblock \emph{Bioinformatics}, 18\penalty0 (1):\penalty0 39--50, 2002.

\bibitem[Nicodemus-Johnson et~al.(2016)Nicodemus-Johnson, Myers, Sakabe,
  Sobreira, Hogarth, Naureckas, Sperling, Solway, White, Nobrega, Nicolae,
  Gilad, and Ober]{nicodemus2016dna}
J.~Nicodemus-Johnson, R.~A. Myers, N.~J. Sakabe, D.~R. Sobreira, D.~K. Hogarth,
  E.~T. Naureckas, A.~I. Sperling, J.~Solway, S.~R. White, M.~A. Nobrega, D.~L.
  Nicolae, Y.~Gilad, and C.~Ober.
\newblock {DNA} methylation in lung cells is associated with asthma endotypes
  and genetic risk.
\newblock \emph{JCI Insight}, 1\penalty0 (20), 2016.

\bibitem[Ortega(1990)]{ortega1990numerical}
J.~M. Ortega.
\newblock \emph{Numerical Analysis: A Second Course}.
\newblock SIAM, 1990.

\bibitem[Parker et~al.(2014)Parker, Bravo, and Leek]{parker2014removing}
H.~S. Parker, H.~C. Bravo, and J.~T. Leek.
\newblock Removing batch effects for prediction problems with frozen surrogate
  variable analysis.
\newblock \emph{PeerJ}, 2:\penalty0 e561, 2014.

\bibitem[Peck and Van~Ness(1982)]{peck1982use}
R.~Peck and J.~Van~Ness.
\newblock The use of shrinkage estimators in linear discriminant analysis.
\newblock \emph{IEEE Transactions on Pattern Analysis and Machine
  Intelligence}, \penalty0 (5):\penalty0 530--537, 1982.

\bibitem[Peres-Neto et~al.(2005)Peres-Neto, Jackson, and Somers]{peres2005many}
P.~R. Peres-Neto, D.~A. Jackson, and K.~M. Somers.
\newblock How many principal components? stopping rules for determining the
  number of non-trivial axes revisited.
\newblock \emph{Computational Statistics \& Data Analysis}, 49\penalty0
  (4):\penalty0 974--997, 2005.

\bibitem[Polley et~al.(2011)Polley, Rose, and Van~der Laan]{polley2011super}
E.~C. Polley, S.~Rose, and M.~J. Van~der Laan.
\newblock Super learning.
\newblock In \emph{Targeted Learning}, pages 43--66. Springer, 2011.

\bibitem[Quay et~al.(1998)Quay, Reed, Samet, and Devlin]{quay1998air}
J.~L. Quay, W.~Reed, J.~Samet, and R.~B. Devlin.
\newblock Air pollution particles induce {IL}-6 gene expression in human airway
  epithelial cells via {NF}-$\kappa$ {B} activation.
\newblock \emph{American Journal of Respiratory Cell and Molecular Biology},
  19\penalty0 (1):\penalty0 98--106, 1998.

\bibitem[Saeys et~al.(2007)Saeys, Inza, and Larra{\~n}aga]{saeys2007review}
Y.~Saeys, I.~Inza, and P.~Larra{\~n}aga.
\newblock A review of feature selection techniques in bioinformatics.
\newblock \emph{Bioinformatics}, 23\penalty0 (19):\penalty0 2507--2517, 2007.

\bibitem[Somineni et~al.(2019)Somineni, Venkateswaran, Kilaru, Marigorta, Mo,
  Okou, Kellermayer, Mondal, Cobb, Walters, Griffiths, Noe, Crandall, Rosh,
  Mack, Heyman, Baker, Stephens, Baldassano, Markowitz, Dubinsky, Cho, Hyams,
  Denson, Gibson, Cutler, Conneely, Smith, and Kugathasan]{somineni2019blood}
H.~K. Somineni, S.~Venkateswaran, V.~Kilaru, U.~M. Marigorta, A.~Mo, D.~T.
  Okou, R.~Kellermayer, K.~Mondal, D.~Cobb, T.~D. Walters, A.~Griffiths,
  J.~Noe, W.~Crandall, J.~Rosh, D.~Mack, M.~Heyman, S.~Baker, M.~Stephens,
  R.~Baldassano, J.~Markowitz, M.~Dubinsky, J.~Cho, J.~Hyams, L.~Denson,
  G.~Gibson, D.~Cutler, K.~Conneely, A.~Smith, and S.~Kugathasan.
\newblock Blood-derived {DNA} methylation signatures of {C}rohn's disease and
  severity of intestinal inflammation.
\newblock \emph{Gastroenterology}, 156\penalty0 (8):\penalty0 2254--2265, 2019.

\bibitem[Sun et~al.(2012)Sun, Zhang, and Owen]{sun2012multiple}
Y.~Sun, N.~Zhang, and A.~Owen.
\newblock Multiple hypothesis testing adjusted for latent variables, with an
  application to the {AGEMAP} gene expression data.
\newblock \emph{Annals of Applied Statistics}, 6\penalty0 (4):\penalty0
  1664--1688, 2012.

\bibitem[Tibshirani et~al.(2002)Tibshirani, Hastie, Narasimhan, and
  Chu]{tibshirani2002diagnosis}
R.~Tibshirani, T.~Hastie, B.~Narasimhan, and G.~Chu.
\newblock Diagnosis of multiple cancer types by shrunken centroids of gene
  expression.
\newblock \emph{Proceedings of the National Academy of Sciences}, 99\penalty0
  (10):\penalty0 6567--6572, 2002.

\bibitem[Virta and Nordhausen(2019)]{virta2019estimating}
J.~Virta and K.~Nordhausen.
\newblock Estimating the number of signals using principal component analysis.
\newblock \emph{Stat}, 8\penalty0 (1):\penalty0 e231, 2019.

\bibitem[Wan et~al.(2012)Wan, Qiu, Baccarelli, Carey, Bacherman, Rennard,
  Agusti, Anderson, Lomas, and DeMeo]{wan2012cigarette}
E.~S. Wan, W.~Qiu, A.~Baccarelli, V.~J. Carey, H.~Bacherman, S.~I. Rennard,
  A.~Agusti, W.~Anderson, D.~A. Lomas, and D.~L. DeMeo.
\newblock Cigarette smoking behaviors and time since quitting are associated
  with differential {DNA} methylation across the human genome.
\newblock \emph{Human Molecular Genetics}, 21\penalty0 (13):\penalty0
  3073--3082, 2012.

\bibitem[Wang et~al.(2017)Wang, Zhao, Hastie, and Owen]{wang2017confounder}
J.~Wang, Q.~Zhao, T.~Hastie, and A.~B. Owen.
\newblock Confounder adjustment in multiple hypothesis testing.
\newblock \emph{Annals of Statistics}, 45\penalty0 (5):\penalty0 1863--1894,
  2017.

\bibitem[Washburn et~al.(2019)Washburn, Wang, Walton, Goedegebuure, Figueroa,
  Van~Horn, Grossman, Remlinger, Madsen, Brown, Srinivasan, Wolf, Berger, Yi,
  Hawkins, Fields, and Hotchkiss]{washburn2019t}
M.~L. Washburn, Z.~Wang, A.~H. Walton, S.~P. Goedegebuure, D.~J. Figueroa,
  S.~Van~Horn, J.~Grossman, K.~Remlinger, H.~Madsen, J.~Brown, R.~Srinivasan,
  A.~Wolf, S.~Berger, V.~Yi, W.~Hawkins, R.~Fields, and R.~Hotchkiss.
\newblock T cell--and monocyte-specific {RNA}-sequencing analysis in septic and
  nonseptic critically ill patients and in patients with cancer.
\newblock \emph{The Journal of Immunology}, 203\penalty0 (7):\penalty0
  1897--1908, 2019.

\bibitem[Witten and Tibshirani(2011)]{witten2011penalized}
D.~M. Witten and R.~Tibshirani.
\newblock Penalized classification using {F}isher's linear discriminant.
\newblock \emph{Journal of the Royal Statistical Society: Series B},
  73\penalty0 (5):\penalty0 753--772, 2011.

\bibitem[Wold(1978)]{wold1978cross}
S.~Wold.
\newblock Cross-validatory estimation of the number of components in factor and
  principal components models.
\newblock \emph{Technometrics}, 20\penalty0 (4):\penalty0 397--405, 1978.

\bibitem[Xu et~al.(2013)Xu, Xu, Yang, Ye, Liu, Wu, Ni, Tan, Cai, Meng, Cai, and
  Du]{xu2013identification}
Y.~Xu, Q.~Xu, L.~Yang, X.~Ye, F.~Liu, F.~Wu, S.~Ni, C.~Tan, G.~Cai, X.~Meng,
  S.~Cai, and X.~Du.
\newblock Identification and validation of a blood-based 18-gene expression
  signature in colorectal cancer.
\newblock \emph{Clinical Cancer Research}, 19\penalty0 (11):\penalty0
  3039--3049, 2013.

\bibitem[Yang et~al.(2010)Yang, Hwa~Yang, B~Zhou, and Y~Zomaya]{yang2010review}
P.~Yang, Y.~Hwa~Yang, B.~B~Zhou, and A.~Y~Zomaya.
\newblock A review of ensemble methods in bioinformatics.
\newblock \emph{Current Bioinformatics}, 5\penalty0 (4):\penalty0 296--308,
  2010.

\bibitem[Zeilinger et~al.(2013)Zeilinger, K{\"u}hnel, Klopp, Baurecht,
  Kleinschmidt, Gieger, Weidinger, Lattka, Adamski, Peters, and
  Strauch]{zeilinger2013tobacco}
S.~Zeilinger, B.~K{\"u}hnel, N.~Klopp, H.~Baurecht, A.~Kleinschmidt, C.~Gieger,
  S.~Weidinger, E.~Lattka, J.~Adamski, A.~Peters, and K.~Strauch.
\newblock Tobacco smoking leads to extensive genome-wide changes in {DNA}
  methylation.
\newblock \emph{PloS One}, 8\penalty0 (5):\penalty0 e63812, 2013.

\bibitem[Zhang et~al.(2012)Zhang, Tibshirani, and
  Davis]{zhang2012classification}
Y.~Zhang, R.~Tibshirani, and R.~Davis.
\newblock Classification of patients from time-course gene expression.
\newblock \emph{Biostatistics}, 14\penalty0 (1):\penalty0 87--98, 2012.

\bibitem[Zheng et~al.(2017)Zheng, Lv, and Lin]{zheng2017nonsparse}
Z.~Zheng, J.~Lv, and W.~Lin.
\newblock Nonsparse learning with latent variables.
\newblock \emph{arXiv:1710.02704}, 2017.

\bibitem[Zou and Hastie(2005)]{zou2005elasticnet}
H.~Zou and T.~Hastie.
\newblock Regularization and variable selection via the elastic net.
\newblock \emph{Journal of the Royal Statistical Society: Series B},
  67\penalty0 (2):\penalty0 301--320, 2005.

\end{thebibliography}

\end{document}